\documentclass[10pt,conference]{IEEEtran}
\usepackage{cite}
\usepackage{amsmath,amssymb,amsfonts}
\usepackage{algorithmic}
\usepackage{graphicx}
\usepackage{textcomp}
\usepackage{xcolor}
\usepackage{wrapfig}
\usepackage{siunitx}

\usepackage{threeparttable}
\usepackage{multirow,multicol}
\usepackage{csvsimple}
\usepackage{tikz}
\usepackage{tcolorbox}
\usepackage{float}
\usepackage{censor}
\usepackage{pifont}
\usepackage{amsmath}
\usepackage{hyperref}
\usepackage{booktabs}
\usepackage{wrapfig}
\usepackage{svg}
\makeatletter
\@namedef{ver@lineno.sty}{9999/12/31}
\@namedef{opt@lineno.sty}{}
\makeatother
\usepackage[frozencache=true,cachedir=minted-cache]{minted} 
\usepackage{balance}
\usepackage{subfig}
\usepackage[hang,flushmargin]{footmisc} 

\newcommand{\eg}{\textit{e.g., }}
\newcommand{\ie}{\textit{i.e., }}

\newcommand{\xmark}{\ding{53}}
\usetikzlibrary{shapes.geometric, arrows, fit}
\usetikzlibrary{positioning}
\usetikzlibrary{decorations.pathreplacing}
\definecolor{tabblue}{HTML}{1f77b4}
\definecolor{taborange}{HTML}{ff7f0e}
\definecolor{tabgreen}{HTML}{2ca02c}
\definecolor{tabred}{HTML}{d62728}
\definecolor{tabpurple}{HTML}{9467bd}
\definecolor{tabbrown}{HTML}{8c564b}
\definecolor{tabpink}{HTML}{e377c2}
\definecolor{tabgray}{HTML}{7f7f7f}
\definecolor{tabolive}{HTML}{bcbd22}
\definecolor{tabcyan}{HTML}{17becf}
\definecolor{rqbeige}{HTML}{faf3f0}
\definecolor{rqpink}{HTML}{f8e8ee}
\definecolor{rqblue}{RGB}{240,240,254}
\definecolor{figblue}{RGB}{203,213,229}
\tikzstyle{framework} = [rectangle, minimum width=2.7cm, minimum height=1cm, text centered, draw=black, fill=rqbeige]
\tikzstyle{input} = [rectangle, minimum width=3.75cm, minimum height=1cm, text centered, draw=black, fill=rqblue]
\tikzstyle{agg} = [rectangle, minimum width=4.2cm, minimum height=1cm, text centered, draw=black, fill=rqpink]
\tikzstyle{agent} = [rectangle, minimum width=2.5cm, minimum height=1cm, text centered, draw=black]
\tikzstyle{env} = [rectangle, minimum width=2.5cm, minimum height=1cm, text centered, draw=black]
\tikzstyle{image} = [rectangle, minimum width=1cm, minimum height=1cm, text centered, draw=black, fill=figblue]
\tikzstyle{fe} = [rectangle, minimum width=1cm, minimum height=2cm, text centered, draw=black]
\tikzstyle{head} = [rectangle, minimum width=0.5cm, minimum height=1cm, text centered, draw=black]
\tikzstyle{arrow} = [thick,->,>=stealth]

\def\BibTeX{{\rm B\kern-.05em{\sc i\kern-.025em b}\kern-.08em
    T\kern-.1667em\lower.7ex\hbox{E}\kern-.125emX}}
\begin{document}

\title{
\hypersetup{hidelinks}
\vspace{-1.5em}
\hfill\href{https://www.acm.org/publications/policies/artifact-review-and-badging-current}{\includegraphics[scale=0.09]{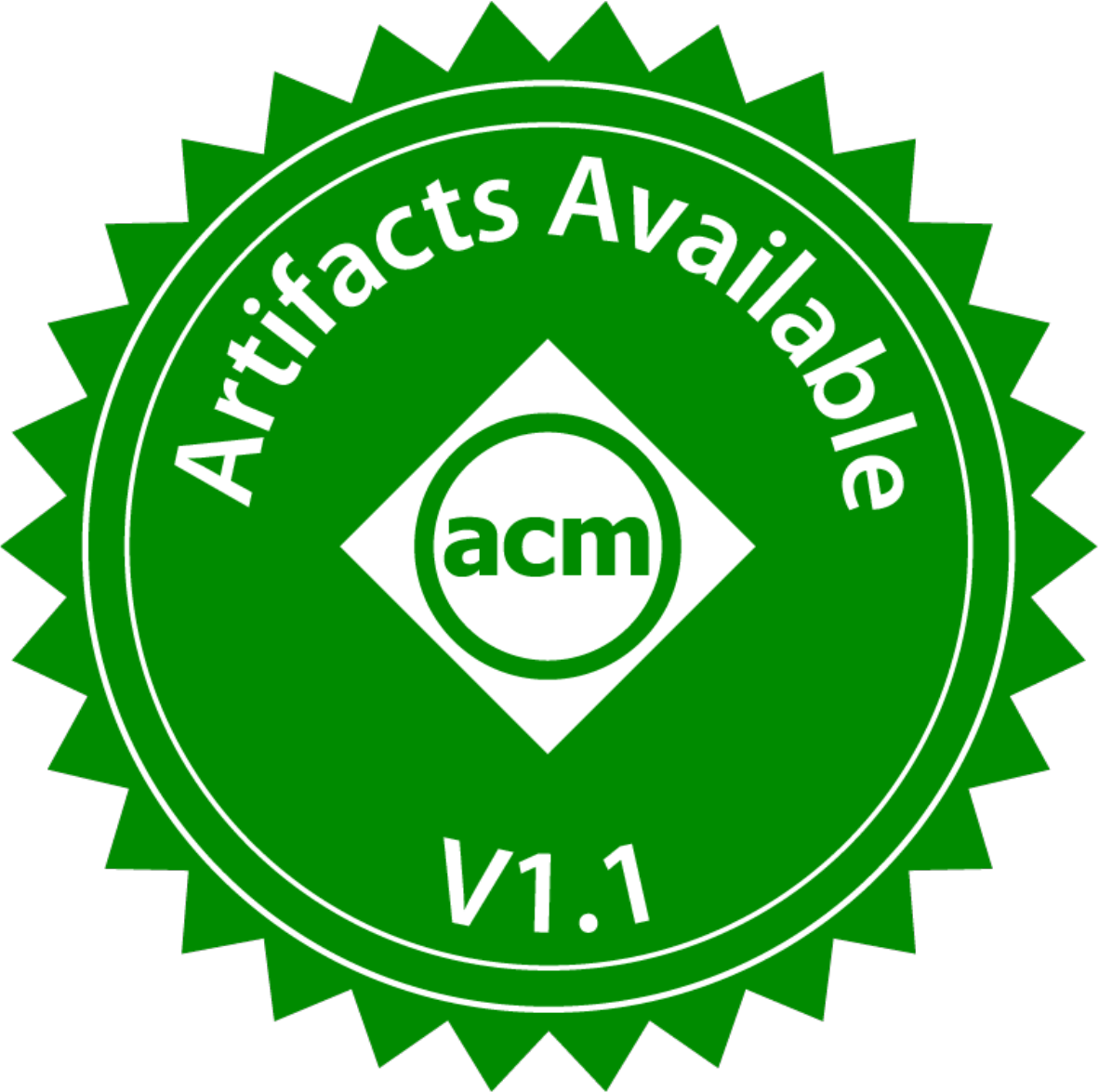}}\hspace{0.5em}\href{https://www.acm.org/publications/policies/artifact-review-and-badging-current}{\includegraphics[scale=0.09]{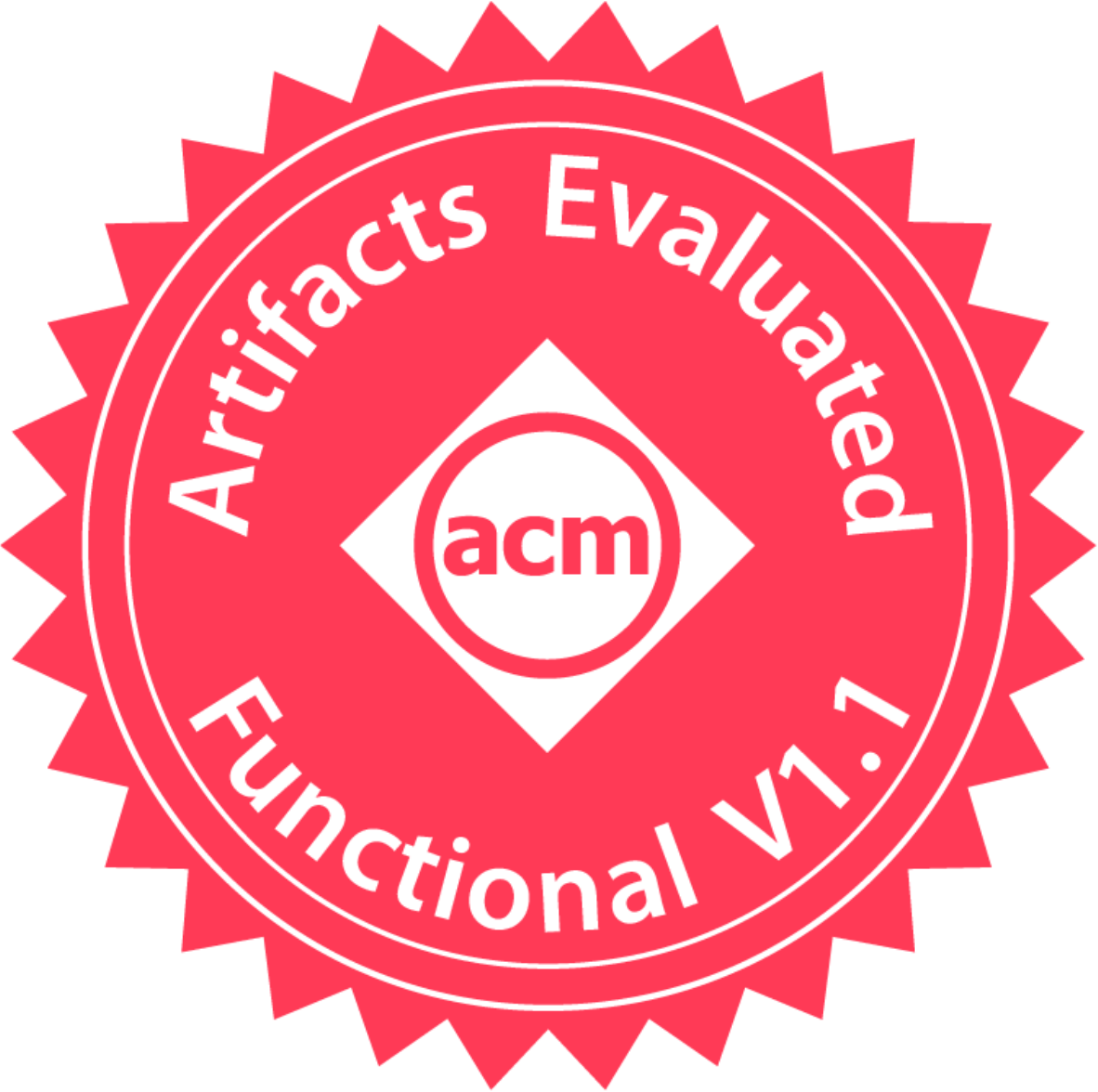}}\hspace{0.5em}\href{https://www.acm.org/publications/policies/artifact-review-and-badging-current}{\includegraphics[scale=0.09]{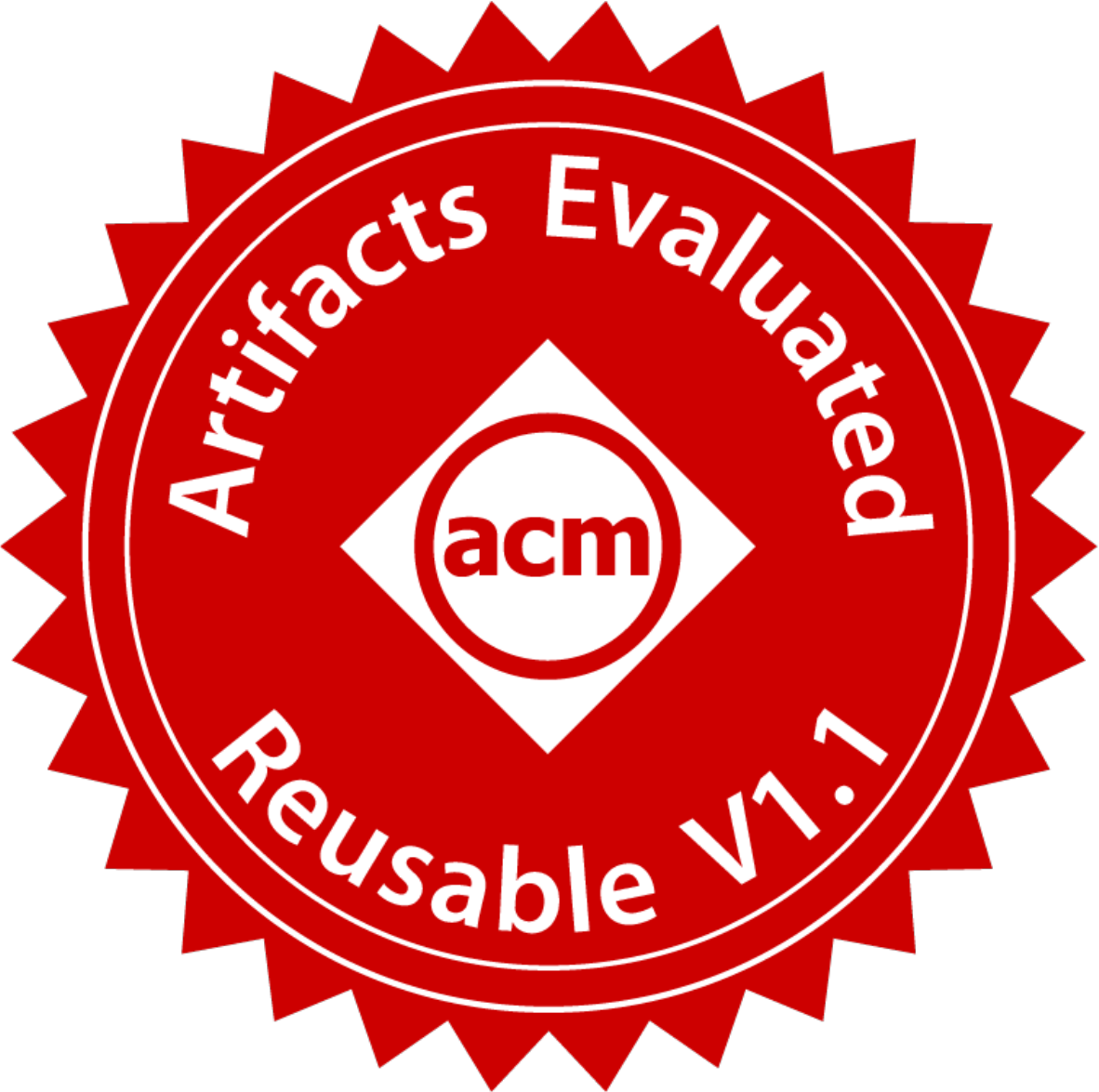}}\\
On the Mistaken Assumption of Interchangeable Deep Reinforcement Learning Implementations
}

\makeatletter
\newcommand{\linebreakand}{%
  \end{@IEEEauthorhalign}
  \hfill\mbox{}\par
  \mbox{}\hfill\begin{@IEEEauthorhalign}
}
\makeatother

\author{\IEEEauthorblockN{Rajdeep Singh Hundal}
\IEEEauthorblockA{National University of Singapore\\
Singapore\\
rajdeep@u.nus.edu}
\and
\IEEEauthorblockN{Yan Xiao\textsuperscript{*\textdagger}}
\IEEEauthorblockA{Sun Yat-sen University\\
Shenzhen, China\\
xiaoy367@mail.sysu.edu.cn}
\and
\IEEEauthorblockN{Xiaochun Cao\textsuperscript{\textdagger}}
\IEEEauthorblockA{Sun Yat-sen University\\
Shenzhen, China\\
caoxiaochun@mail.sysu.edu.cn}
\linebreakand
\IEEEauthorblockN{Jin Song Dong}
\IEEEauthorblockA{National University of Singapore\\
Singapore\\
dcsdjs@nus.edu.sg}
\and
\IEEEauthorblockN{Manuel Rigger}
\IEEEauthorblockA{National University of Singapore\\
Singapore\\
rigger@nus.edu.sg}
}

\maketitle
\begingroup\renewcommand\thefootnote{*}
\footnotetext{Corresponding author.}
\endgroup
\begingroup\renewcommand\thefootnote{\textdagger}
\footnotetext{Shenzhen Campus of Sun Yat-sen University.}
\endgroup
\begingroup\renewcommand\thefootnote{}
\footnotetext{This work was published in the 2025 IEEE/ACM 47th International Conference on Software Engineering (ICSE)~\cite{hundal2025mistaken} and is available in \href{https://doi.org/10.1109/ICSE55347.2025.00222}{IEEE Xplore}.}
\endgroup
\begingroup\renewcommand\thefootnote{}
\footnotetext{\textcopyright\text{ }2025 IEEE. Personal use of this material is permitted. Permission from IEEE must be obtained for all other uses, in any current or future media, including reprinting/republishing this material for advertising or promotional purposes, creating new collective works, for resale or redistribution to servers or lists, or reuse of any copyrighted component of this work in other works.}
\endgroup

\begin{abstract}
\emph{Deep Reinforcement Learning} (DRL) is a paradigm of artificial intelligence where an \emph{agent} uses a neural network to learn which actions to take in a given \emph{environment}.
DRL has recently gained traction from being able to solve complex environments like driving simulators, 3D robotic control, and multiplayer-online-battle-arena video games.
Numerous \emph{implementations} of the state-of-the-art algorithms responsible for training these agents, like the \emph{Deep Q-Network} (DQN) and \emph{Proximal Policy Optimization} (PPO) algorithms, currently exist.
However, studies make the mistake of assuming implementations of the same algorithm to be consistent and thus, \emph{interchangeable}.
In this paper, through a \emph{differential testing} lens, we present the results of studying the extent of implementation inconsistencies, their effect on the implementations' performance, as well as their impact on the conclusions of prior studies under the assumption of interchangeable implementations.
The outcomes of our differential tests showed significant discrepancies between the tested algorithm implementations, indicating that they are \textit{not} interchangeable. 
In particular, out of the five PPO implementations tested on 56 games, three implementations achieved superhuman performance for 50\% of their total trials while the other two implementations only achieved superhuman performance for less than 15\% of their total trials.
Furthermore, the performance among the high-performing PPO implementations was found to differ significantly in nine games.
As part of a meticulous manual analysis of the implementations' source code, we analyzed implementation discrepancies and determined that code-level inconsistencies primarily caused these discrepancies.
Lastly, we replicated a study and showed that this assumption of implementation interchangeability was sufficient to \emph{flip} experiment outcomes.
Therefore, this calls for a shift in how implementations are being used.
In addition, we recommend for (1) replicability studies for studies mistakenly assuming implementation interchangeability, (2) DRL researchers and practitioners to adopt the differential testing methodology proposed in this paper to combat implementation inconsistencies, and (3) the use of large environment suites.
\end{abstract}

\begin{IEEEkeywords}
reinforcement learning, differential testing
\end{IEEEkeywords}
\section{Introduction}
\emph{Deep Learning} (DL) and \emph{Deep Reinforcement Learning} (DRL) are popular paradigms of \emph{Artificial Intelligence} (AI) that use neural networks to solve a problem. 
Different from DL, where the dataset is fixed and re-used throughout the training process, DRL allows for online learning in a controlled \emph{environment} where the dataset is not fixed but rather procured from the environment on-the-fly---the very reason it is chosen over DL in some scenarios such as video games and autonomous driving~\cite{kiran2021deep,abdulazeez2024review,eldahshan2022deep}.
More formally, DRL is a paradigm of AI where a program, or more specifically, an \emph{agent}, learns the optimal action to take in an environment after iterating multiple times through it. 
DRL has proven to be extremely effective in games with recent advances in the field; for example, DeepMind's AlphaZero~\cite{doi:10.1126/science.aar6404} became the first algorithm to beat a world-champion computer program at Chess, Shogi, and Go and OpenAI's OpenAI Five~\cite{openai2019dota} became the first algorithm to defeat the human world champions at the popular online video game Dota 2. 
Furthermore, DRL's applications span beyond those of games, from time-dependent systems like autonomous vehicles~\cite{li2022metadrive} and stock trading~\cite{liu2021finrl} to precision-focused systems like 3D robotic control~\cite{mnih2016asynchronous,schulman2017proximal}.

\sloppy{Similar to DL \emph{libraries} like TensorFlow and PyTorch, numerous DRL libraries providing algorithm \emph{implementations} have been created.}
With more than 10K GitHub stars, RLlib~\cite{liang2018rllib}, Baselines~\cite{baselines}, and Dopamine~\cite{castro18dopamine} are among the most popular DRL libraries. 
RLlib from Ray specialises in the distributed training of DRL algorithms, at a scalable level. 
Baselines from OpenAI and Dopamine from Google are research-focused libraries, built towards the quick replication and refinement of DRL algorithms. 

In DRL studies, it is common to conduct comparisons between different algorithms~\cite{mnih2016asynchronous,schulman2017proximal,schulman2015trust,hessel2018rainbow}.
Moreover, as multiple implementations of an algorithm currently exist, some comparative studies use unoriginal implementations~\cite{zeng2024offline,islam2017reproducibility,raghunath2022reinforcement,wolk2022beyond}.
This is because researchers---as well as practitioners---assume that an algorithm would perform equally well across its implementations and thus, use them \emph{interchangeably}.
To demonstrate this, we systematically conducted a literature review which included research papers from two popular AI conferences and identified 23 research papers making this assumption, with publishing dates ranging from 2017 to 2024.

DRL algorithms are \textit{not} guaranteed to be consistent. On the contrary, they have a \textit{high} risk of inconsistencies. 
Firstly, this is due to the large amount of \emph{hyperparameters}---tunable values that guide and control the learning process.
Being a combination of both DL and \emph{Reinforcement Learning} (RL), DRL inherits hyperparameters from both of them. 
Secondly, because of the large amount of hyperparameters, DRL libraries run a higher risk of (1) making mistakes, and (2) improving or worsening algorithms via seemingly minor code-level implementation choices~\cite{Engstrom2020Implementation}.
This results in DRL libraries implementing their own \textit{flavour} of an algorithm, leading to similar but distinct implementations of the same algorithm in the DRL domain. 

We conducted a pilot study to attain an \textit{initial} assessment on the extent and effects of the aforementioned implementation inconsistencies. 
In particular, we compared five implementations of the \emph{Deep Q-Network} (DQN) algorithm~\cite{mnih2015human} and observed performance discrepancies as well as code-level inconsistencies.
If a mature algorithm like DQN, which has already been extensively studied, standardized, and built upon, exhibits such disparity across some of the most popular implementations, it raises concerns about (1) the level of inconsistency and potential issues with the implementations of less mature algorithms, and (2) the validity of existing research that assume interchangeable algorithm implementations.
Varying efficacies of the \emph{same} algorithm across its implementations might render comparisons from studies that assume interchangeability invalid.
This could have wide-ranging effects, as the conclusions of many studies might need to be re-examined.

In this paper, we assess the extent of implementation inconsistencies in the context of DRL and how they affect research under the assumption that implementations are consistent and interchangeable.
At a high-level, our study relies on \emph{differential testing}, which is a well-known software testing methodology~\cite{mckeeman1998differential}. 
In particular, we conducted a large-scale study (of approximately 10K GPU hours) to compare the efficacy of multiple implementations of the same algorithm to identify potential inconsistencies. 
Using differential testing, we answer the following research questions:
\begin{itemize}
    \item \textbf{RQ1: How prevalent are discrepancies between implementations of the same algorithm?} 
    Varying efficacies of an algorithm across its implementations would produce untrustworthy results and conclusions from studies that assume implementation interchangeability---using alternate implementations over the original.
    Thus, it is important to study the extent of these variances first before assuming that implementations are interchangeable.
    We answered RQ1 by using differential testing and state-of-the-art DRL comparison techniques to assess different implementations of the \emph{Proximal Policy Optimization} (PPO) algorithm~\cite{schulman2017proximal} in terms of efficacy (\ie mean reward), with statistical guarantees.
    \item \textbf{RQ2: Why do implementation discrepancies occur?}
    Understanding why implementation discrepancies occur is vital to creating effective solutions. Thus, we answered RQ2 by investigating the root cause of the discrepancies found in RQ1. In particular, similar to the pilot study, we inspected the implementations' source code for inconsistencies that accounted for the discrepancies found.
    \item \textbf{RQ3: Can the assumption of interchangeable implementations alter the outcomes of an experiment?}
    There would be cause for concern if using a different algorithm implementation was significant enough to \textit{flip} experiment outcomes, as studies which assumed implementation interchangeability might then need to be re-examined.
    Thus, to determine if this was possible, we answered RQ3 by replicating a study that assumed implementation interchangeability~\cite{islam2017reproducibility}, but with a different \emph{Deep Deterministic Policy Gradient} (DDPG) implementation~\cite{lillicrap2015continuous}, and compared the outcomes.
\end{itemize}

Our experiments showed that DRL implementations were \textit{mistaken} to be interchangeable, leading to significant alterations in experimental outcomes. 
In particular, out of the five PPO implementations tested on 56 environments, three implementations consistently outperformed the other two implementations in all comparison techniques used.
Moreover, we statistically show that even the high-performing PPO implementations differed significantly among themselves in nine environments. 
When investigating the root cause of these discrepancies, we found multiple code-level inconsistencies that accounted for the discrepancies found.
Lastly, our experiments showed that this assumption of implementation interchangeability was also significant enough to alter experiment outcomes by \textit{flipping} two out of the three outcomes tested from the replicated study.

Thus, we recommend for (1) replicability studies for studies mistakenly assuming implementation interchangeability, (2) DRL researchers and practitioners to adopt the differential testing methodology proposed in this paper to combat implementation inconsistencies, and (3) the use of large environment suites.
Furthermore, a key implication for researchers studying software testing is that, despite DRL's stochasticity and non-deterministic output, differential testing still can be applied. Therefore, the techniques used in this paper are potentially applicable to non-AI stochastic systems that can benefit from differential testing.
Our source code is publicly available and can be found at \href{https://doi.org/10.5281/zenodo.14249024}{https://doi.org/10.5281/zenodo.14249024}.

\section{Background}

In this section, we discuss the necessary terminology used in this study. 

\paragraph{Environment and agent}
RL has two basic components, the agent and the environment. 
The agent can be viewed as the main processing component and the environment represents the domain the agent tries to solve.
At any arbitrary timestep $t$, the environment passes the current state $S_t$ and reward $R_t$ to the agent, which in turn executes an action $A_t$ in the environment. 
Agents use various \emph{algorithms} to determine the best possible action $A_t$ at state $S_t$. 
In our study, we primarily considered \emph{Q-learning} (QL)~\cite{watkins1989learning} and \emph{Policy Gradient} (PG)~\cite{sutton1999policy} based algorithms as they were commonly implemented by libraries.

\begin{figure}[t]
    \centering
    \includegraphics[width=\linewidth]{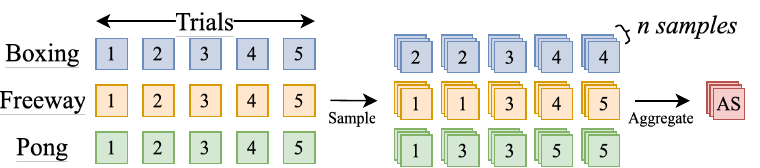}

    \caption{Applying SBCI to all trials from a configuration.}
    \label{fig:sbci}
\end{figure}

\paragraph{Stratified bootstrap confidence intervals}
Stratified Bootstrap Confidence Intervals (SBCI)~\cite{agarwal2021deep} is a systematic and standardized way of reporting performance-based metrics, like an agent's mean reward, in DRL.
SBCI involves bootstrapping the results of a few similarly configured trials to create a distribution that represents the true result distribution for that particular configuration more accurately, as illustrated in~\autoref{fig:sbci}.
All trials (left) are (1) stratified bootstrapped (sampling with replacement, with equal proportions per environment) to form a \emph{sample} (middle), and (2) subsequently aggregated according to a metric (\eg mean) to form an \emph{aggregated sample} (AS).
After \emph{multiple} aggregated samples are accumulated, a distribution is then formed with confidence guarantees.
SBCI is particularly useful when testing an algorithm over a large suite of environments since it significantly reduces the number of trials needed per environment for reliable estimates.

\section{Motivation}
In this section, we discuss the literature review and the pilot study that motivated this paper.

\subsection{Literature Review}
Although we were aware of studies that made the assumption of interchangeable algorithm implementations~\cite{zeng2024offline,islam2017reproducibility,raghunath2022reinforcement,wolk2022beyond}, we decided to conduct a \textit{systematic} and \textit{reproducible} literature review to properly justify this assumption.

\paragraph{Methodology}
We reviewed recently accepted papers from two popular AI conferences---the conference on \emph{Neural Information Processing Systems} (NeurIPS) and the \emph{International Conference on Machine Learning} (ICML), both of the year 2023.
Since both conferences combined had more than 5K accepted papers (3,584 for NeurIPS and 1,865 for ICML), we used the conference data provided by Paper Copilot~\cite{papercopilot} and a Python library named PDFPlumber~\cite{Singer-Vine_pdfplumber_2024} to automatically download the accepted papers and filter them by searching for keywords.
We selected these keywords based on \textit{two assumption scenarios} that primarily occurred in the studies we were already aware of; (1) when studies use an implementation that was not the original~\cite{zeng2024offline,islam2017reproducibility} or (2) when studies alternate between two different implementations of the \textit{same} algorithm in their experiments~\cite{raghunath2022reinforcement,wolk2022beyond}, with the latter being a manifestation of the former.
Thus, to narrow down our literature to review, we searched for papers which had the key phrase “Reinforcement Learning" in their title as well as references to at least two popular DRL libraries (see~\autoref{tab:github} for the list of popular DRL libraries).
The keywords used to search for popular DRL libraries were “RLlib, Baselines3, Tianshou, CleanRL, and Baselines". 
For completeness, all searches were done in a case-insensitive manner.

\paragraph{Analysis}
In total, we identified ten papers from NeurIPS, six papers from ICML, and seven papers from other conferences under the assumption of interchangeable algorithm implementations. 
Examples of studies conforming to the first aforementioned scenario are the works by Beukman et al.~\cite{beukman2024dynamics}, Chiappa et al.~\cite{chiappa2024latent}, and Gerstgrasser et al.~\cite{gerstgrasser2024selectively}, where alternate implementations were used over the original for the SAC~\cite{haarnoja2018soft}, PPO, and DQN algorithms respectively.
Examples of studies conforming to the second aforementioned scenario are the works by Raghunath et al.~\cite{raghunath2022reinforcement} and Wolk et al.~\cite{wolk2022beyond}.
Raghunath et al.~\cite{raghunath2022reinforcement} integrated their study with the SAC algorithm in a single-agent setting as well as a multi-agent setting to demonstrate effectiveness. However, Raghunath et al. used different SAC implementations for both settings, assuming that they were consistent and thus, interchangeable. Given that their experiments show marginal differences between both single-agent and multi-agent settings~\cite[Figure 2]{raghunath2022reinforcement}, if the SAC implementations were \textit{not} interchangeable, a change in SAC implementation could easily \textit{flip} the experimental outcomes.
Similarly, Wolk et al.~\cite{wolk2022beyond} integrated their study with the PPO algorithm in multiple settings. However, Wolk et al. used different PPO implementations for some settings, assuming they were interchangeable. This potentially invalidates their comparisons, especially for those with marginal differences~\cite[Table 2]{wolk2022beyond}. 

\paragraph{Key takeaways}
With numerous studies found to be under the assumption of interchangeable algorithm implementations, the notion that researchers---as well as practitioners---assume interchangeability is well-justified and potentially warrants study re-examinations if implementations were found to \textit{not} be interchangeable.

\begin{figure*}[t]
    \centering
    \includegraphics[width=\textwidth]{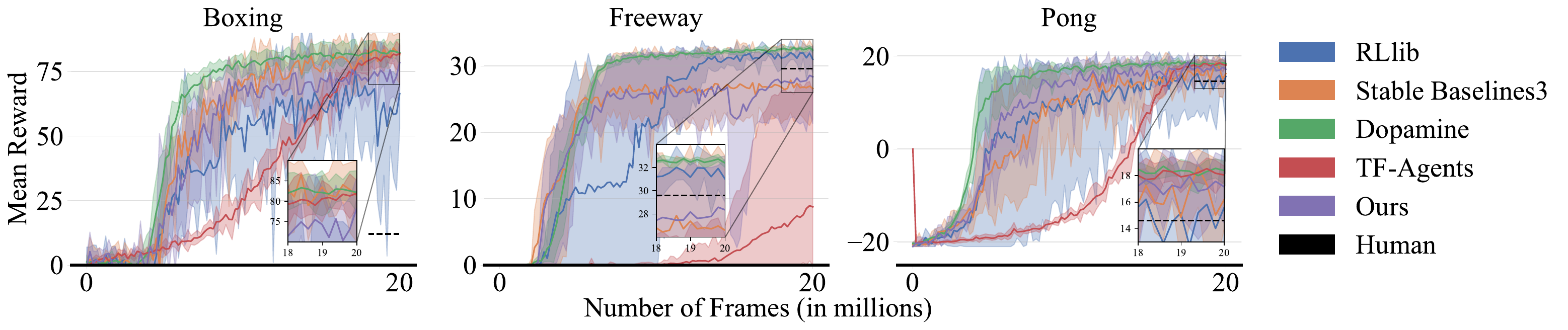}
    \caption{Training curves from five DQN implementations where the y-axis represents the mean in-game reward while the x-axis represents the number of in-game frames that have passed. Five agents were trained for each (implementation, game) permutation and the training curves were aggregated to display the mean, minimum, and maximum within the shaded regions. The mean reward attained by a professional human tester is also shown in black to gauge superhuman or subhuman capabilities.}
    \label{fig:teaser}
\end{figure*}
\begin{figure}[t]
    \centering
    \subfloat{{\includegraphics[width=0.24\textwidth]{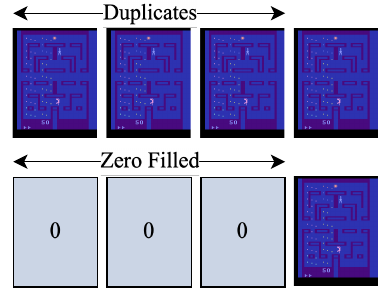}}}
    \hfill
    \subfloat{{\includegraphics[width=0.245\textwidth]{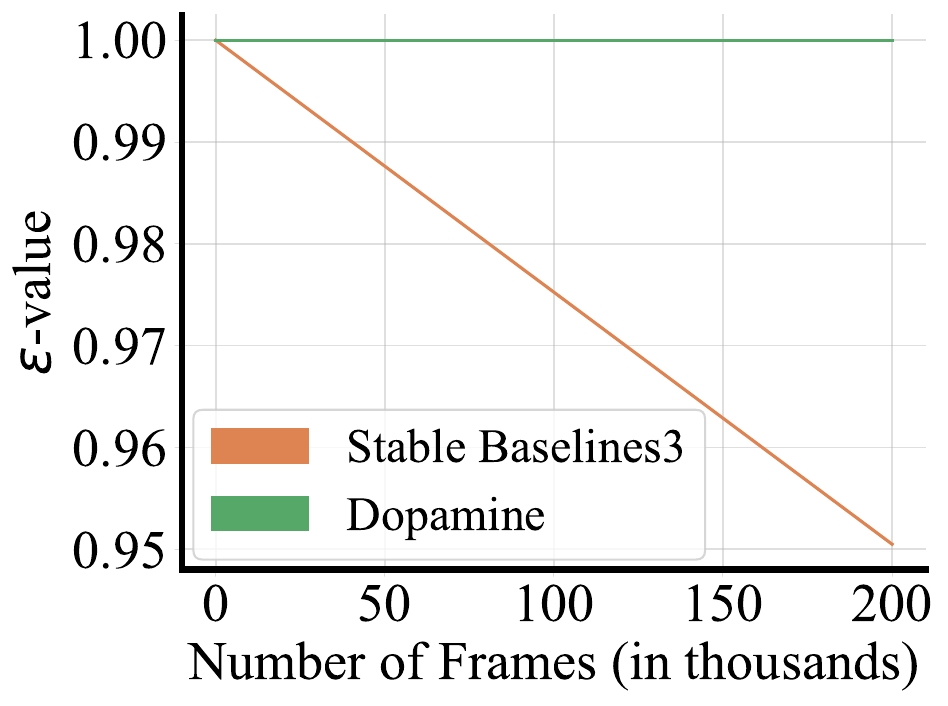}}}
    \caption{Frame stacking approaches taken by Stable Baselines3 (top-left) and Dopamine (bottom-left), as well as a comparison of their implementation of the $\epsilon$-greedy policy (right).}
    \label{fig:eps}
\end{figure}
\subsection{Pilot Study}
We conducted a pilot study to attain an \textit{initial} assessment on the extent and effects of implementation inconsistencies before conducting a more thorough comparison on a larger scale as DRL experiments can often be computationally expensive. 

\paragraph{Methodology} We trained agents from five DQN implementations (with the same hyperparameters) to solve three games from the Atari 2600 benchmark~\cite{bellemare13arcade}, as illustrated in~\autoref{fig:teaser}.
Since there were installation issues with the original DQN implementation, and the DQN's authors subsequently vouched for Dopamine's implementation as an accurate alternative in a GitHub issue, we used Dopamine's DQN implementation as the baseline.\footnote{All links to issues and pull requests are included in the source code.}
Moreover, we also included our own DQN implementation (\ie Ours).

\paragraph{Analysis}
We observed stark discrepancies in the training curves, with the majority of implementations performing suboptimally when compared to Dopamine, with either low final mean rewards or high variance among their trials (\ie elongated shaded regions).
To account for these discrepancies, we investigated code-level inconsistencies.
Firstly, we found that Stable Baselines3~\cite{stable-baselines3} and Dopamine both differ in how they form the stack of frames as input to the Q-network, with the former duplicating frames to account for missing frames and the latter filling the missing frames with zeros instead, as illustrated in~\autoref{fig:eps} (left).
Secondly and more importantly, we noticed that they also differ in how they decay certain hyperparameters, like the $\epsilon$-value for the $\epsilon$-greedy policy---a strategy where the $\epsilon$-value (decayed with time) encourages an agent's exploration in an environment. 
As shown in~\autoref{fig:eps} (right), Stable Baselines3 starts the decay immediately, regardless of how many frames have passed, while Dopamine waits and only starts the decay much later on.
As this directly impacts how an agent behaves, it could explain the discrepancy, in terms of curve steepness, between the two implementations.

\paragraph{Key takeaways}
The discovery of efficacy discrepancies as well as major code-level inconsistencies among implementations of an already mature algorithm suggests the conjecture that implementations are \textit{not} interchangeable and warrants a study of larger scale, with less mature algorithms and more environments.

\section{Methodology}
In this section, we discuss the approach we adopted to address the three RQs. In particular, (1) how we selected, compared, and debugged implementations, (2) how we replicated a study, and (3) the experimental settings.

\begin{table}[t]
\centering
\begin{threeparttable}
\caption{DRL Libraries}
\label{tab:github}
\begin{tabular}{@{}lcccrr@{}}
\toprule
DRL Library & $\frac{1}{2}$\tnote{1} & DQN & PPO & Stars (K) & Last Commit \\
\midrule
\textbf{RLlib} & \checkmark & \checkmark & \checkmark & 31.6\tnote{2} & 2024 \\
Dopamine & \checkmark & \checkmark & \xmark & 10.4 & 2024 \\
\textbf{Stable Baselines3} & \checkmark & \checkmark & \checkmark & 8.2 & 2024 \\
\textbf{Tianshou} & \checkmark & \checkmark & \checkmark & 7.5 & 2024 \\
\textbf{CleanRL} & \checkmark & \checkmark & \checkmark & 4.6 & 2024 \\
OpenSpiel & \xmark & \checkmark & \checkmark & 4.1 & 2024 \\
ReAgent & \checkmark & \checkmark & \checkmark & 3.5 & 2024 \\
ElegantRL & \checkmark & \checkmark & \checkmark & 3.5 & 2024 \\
Acme & \checkmark & \checkmark & \checkmark & 3.4 & 2024 \\
Tensorforce\tnote{3} & \checkmark & \checkmark & \checkmark & 3.3 & 2024 \\
TF-Agents & \checkmark & \checkmark & \checkmark & 2.7 & 2024 \\
TorchRL & \checkmark & \checkmark & \checkmark & 2.0 & 2024 \\
Coach\tnote{3} & \checkmark & \checkmark & \checkmark & 2.3 & 2022 \\ 
Tonic RL & \checkmark & \xmark & \checkmark & 0.39 & 2021 \\
Catalyst-RL & \checkmark & \checkmark & \checkmark & 0.046 & 2021 \\
\textbf{Baselines} & \checkmark & \checkmark & \checkmark & 15.4 & 2020 \\
Spinning Up\tnote{4} & \checkmark & \xmark & \checkmark & 9.7 & 2020 \\
Keras-RL & \checkmark & \checkmark & \xmark & 5.5 & 2019 \\
\bottomrule
\end{tabular}
\begin{tablenotes}
\item[1] At least half of the algorithms implemented are QL or PG-based.
\item[2] With respect to the entire Ray codebase and not just RLlib.
\item[3] Will no longer be maintained.
\item[4] Does not support GPU computations.
\end{tablenotes}
\end{threeparttable}
\end{table}

\subsection{Selecting Algorithms and Implementations}
Due to significant hardware requirements, testing and comparing all algorithms with their respective implementations would have been infeasible.
Thus, in order to assess the prevalence of discrepancies between different implementations of the same algorithm in RQ1, we first determined which algorithms and implementations were representative of the DRL domain.
\autoref{tab:github} depicts a list of popular DRL libraries, sorted firstly by last commit date, and secondly by GitHub stars.
Furthermore, we investigated the type of algorithms that were commonly implemented and determined that (1) QL or PG-based algorithms were the most common, and (2) DQN and PPO were the most common QL and PG-based algorithms respectively.
Subsequently, we narrowed our scope to selecting just one algorithm, as it then allowed us to conduct a \emph{large-scale} comparison across more implementations with a large environment suite.

We chose to test and compare PPO implementations over DQN implementations for RQ1 due to their low memory costs.
Memory was a primary concern because it determines how many trials we were able to run concurrently on our GPU servers.
Since the memory costs for DQN are significantly higher than that of PPO (\eg 1M states vs 1K states in the Atari 2600 benchmark~\cite{mnih2015human,schulman2017proximal}), we chose to test PPO implementations instead.
To put this in a better perspective, we could run eight concurrent PPO trials compared to two concurrent DQN trials on our GPU servers.
Moreover, we limited our selection of PPO implementations to just five implementations, so that we could incorporate a larger suite of environments in our tests.
Thus, we selected the implementation by Baselines since it was the original PPO implementation and subsequently selected implementations from the four most popular actively maintained libraries for RQ1, to wit, RLlib, Stable Baselines3, Tianshou~\cite{tianshou}, and CleanRL~\cite{huang2022cleanrl}.
It would be a strong indication that libraries are inconsistent with their implementations of other algorithms if they were already inconsistent with a widely used algorithm, like PPO.

\subsection{Comparing DRL Performance}
Differential testing compares the outputs of two systems when given the same input.
Should the outputs be the same, the systems are considered to be correct for the given input.
Conversely, should the outputs be different, the systems are considered to have differing behaviour.
Differential testing is primarily used as a black-box testing technique and thus, widely applicable for systems with neural networks~\cite{deng2022fuzzing,guo2020audee,wei2022free,duo,fuzzgpt}.
However, the same differential testing methodology used by the aforementioned DL studies cannot be directly applied to DRL without first addressing the domain's stochasticity.

\paragraph{Differential testing in DRL} The core challenge of applying differential testing to DRL systems is the inherent stochasticity present in the systems.
A given input might not consistently produce the same output, as observed in the pilot study, in~\autoref{fig:teaser}.
We address this uncertainty by using SBCI to attain accurate and reliable estimates.
In particular, when testing PPO implementations in RQ1, the \emph{inputs} were PPO's hyperparameters---an algorithm's configuration, manually set to be consistent with the original publication, across all tested implementations. To compare fairly with other PPO implementations, utilizing SBCI, the \emph{outputs} of all trials from a PPO implementation were stratified bootstrapped with respect to multiple \emph{aggregation metrics}, to evaluate different aspects of efficacy.
This results in \emph{point estimates} with \emph{confidence bands} for all aggregation metrics used, for all PPO implementations tested, allowing for a more accurate and reliable comparison.

\paragraph{Environments} Since an agent's performance is dependent on the difficulty of an environment~\cite{bellemare13arcade}, to further increase the fairness of our comparisons in RQ1, we used a large and diverse suite of environments from the Atari 2600 benchmark (56 environments) when testing the PPO implementations---the same environments commonly used to benchmark new algorithms~\cite{mnih2016asynchronous,schulman2017proximal,hessel2018rainbow,mnih2015human,van2016deep,wang2016dueling}.
This was done to ensure that all implementations were tested in as many situations as possible to better compare their agents' generalization capabilities---to perform in any environment.

\paragraph{Human normalized score}
Prior to the application of any statistical techniques post-training, it is common to first normalize the outputs (\ie mean rewards) of trials from the Atari 2600 benchmark suite with respect to the mean rewards attained by a professional human tester, so as to properly gauge superhuman performance in games~\cite{hessel2018rainbow,mnih2015human,van2016deep,wang2016dueling}.
Thus, to obtain a single human normalized score representing the efficacy of a PPO trial in RQ1, we first took the mean reward attained from the last 100 training episodes instead of just the reward from the last training episode for a more accurate estimate of an agent's efficacy~\cite{machado2018revisiting,schulman2017proximal}. 
Subsequently, we applied min-max normalization to the mean reward,
\begin{equation}
    \label{eq:normalization}
    Score = \frac{MeanReward_{100} - RandomPlay}{HumanPlay - RandomPlay}
\end{equation}
where min and max represent the reward attained from random and human play respectively, referenced from a previous study~\cite{wang2016dueling}. 
The agent is superhuman if $Score > 1$.

\paragraph{Statistical techniques}
In this study, it is essential that we draw statistically-sound conclusions about any potential discrepancies among the implementations, particularly when dealing with stochastic algorithms.
To this end, aside from statistical tests like ANOVA~\cite{girden1992anova}, we used a variety of state-of-the-art techniques specifically tailored for DRL.

Firstly, we incorporated an environment-wise one-way ANOVA to compare the effect of PPO implementation on $MeanReward_{100}$. 
With a null hypothesis of equal implementation means, rejecting it (\ie p-value $<$ 0.05) translates to a statistically significant difference in means. 
However, (1) a one-way ANOVA does not pinpoint outliers, (2) with large amounts of data, minor effects can easily cause statistically significant differences, and (3) statistical insignificance does not imply absence of effect~\cite{greenland2016statistical,agarwal2021deep}. Thus, we decided to include additional techniques for broader perspectives.

Subsequently, to further compare the PPO implementations, we applied SBCI to their scores with respect to two different aggregation metrics~\cite{agarwal2021deep}; (1) fraction of trials with $Score > \tau$, and (2) probability of improvement (POI).
The first metric results in a \emph{performance profile} across all of an implementation's trials, to gauge the number of trials that achieved superhuman game performance (\ie when $\tau = 1)$.
The second metric uses more direct, one-to-one \emph{pairwise comparisons} with the Mann-Whitney U-statistic~\cite{mann1947test}, that is,
\begin{equation}
    \label{eq:mwu}
    \begin{split}
       P(X_m > Y_m) = \frac{1}{N^2} \sum_{i=1}^{N}\sum_{j=1}^{N} S(x_{m,i},y_{m,j}),\\ 
       where \;\; S(x,y) = 
       \begin{cases}
           1, & \text{if}\;\; y<x, \\
           \frac{1}{2}, & \text{if}\;\; y=x, \\
           0, & \text{if}\;\; y>x.
       \end{cases}
    \end{split}
\end{equation}
Here, we directly compute the POI of implementation $X$ over implementation $Y$ with $N$ trials each, on environment $m$.
The primary assumption is that for $X$ to be better than $Y$, $X$ has to outperform $Y$ \emph{sufficiently often}.
The final POI of $X$ over $Y$, $P(X>Y)$, is subsequently attained by calculating the POI for all environments $\frac{1}{M}\sum_{m=1}^{M} P(X_m > Y_m)$.
Post-SBCI, the point estimates and confidence bands attained from this metric are further analyzed with the Neyman-Pearson testing criterion~\cite{bouthillier2021accounting,neyman1928use} for statistical significance and meaningfulness.
Statistical significance where $P(X>Y)>0.5 \;\wedge\; 0.5 \notin CI$ rules out the effect of noise on the point estimates. On the other hand, statistical meaningfulness where $CI_{upper}>0.75$ ensures that $X$ outperforms $Y$ often enough.
X is considered to be \emph{better} than Y if both aforementioned criteria are met.
Thus, the POI metric directly measures the likelihood that an implementation $X$ outperforms implementation $Y$ on a random environment, complementing the performance profiles.

\subsection{Debugging Discrepancies}
To locate the root cause of discrepancies found in RQ1 for RQ2, we adopted a \emph{best-effort} debugging approach similar to that of the pilot study. 
Moreover, we focused on the high-performing PPO implementations as they likely differed from code-level inconsistencies rather than critical bugs.
In particular, we manually inspected the PPO implementations' source codes to identify any inconsistencies that could account for the discrepancies found.
Upon finding any inconsistencies, we (1) corrected them to be consistent across implementations, (2) re-tested the implementations, and (3) re-compared the implementations to determine if the discrepancies were still present. 
Although time-consuming as this required us to build an in-depth understanding of the source codes, it was necessary in order to ascertain that the inconsistencies were indeed the cause of the discrepancies found.
Lastly, we also informed the implementations' developers, via GitHub issues, of the unresolved discrepancies, as a final attempt to address them.

\subsection{Replicating a Study}
In order to determine if the assumption of interchangeable implementations could alter the outcomes of an experiment in RQ3, we replicated experiments from a study that made the assumption.
Islam et al.~\cite{islam2017reproducibility} opted to use an alternate implementation of the DDPG algorithm over the original~\cite{lillicrap2015continuous} and recommended ideal hyperparameter values for the DDPG algorithm, assuming that the DDPG implementations were interchangeable.
We chose this specific study because (1) many works based on the study's results, as its more than 300 citations show, (2) it adhered to the current best practices by providing detailed descriptions of the hyperparameter values and implementations used, and (3) it was simple.
In particular, the study investigated the best value for a few commonly used hyperparameters, in terms of efficacy.
The study tested different values for each hyperparameter investigated with DDPG on the HalfCheetah environment from the MuJoCo benchmark suite~\cite{todorov2012mujoco}.
We replicated the experiments involving the network architecture, reward scale, and batch size hyperparameters, but with a different DDPG implementation---Stable Baselines3's DDPG implementation---to investigate if using a different implementation was significant enough to \textit{flip} experiment outcomes.
Furthermore, we used the same differential testing methodology as in RQ1 (and RQ2), for accurate and reliable comparisons.

\subsection{Experimental Settings}
We conducted our experiments concurrently on three GPU servers running the Ubuntu LTS operating system.
The first server had three NVIDIA RTX A4000 GPUs with an AMD Ryzen Threadripper 3970X CPU.
The second server had four NVIDIA RTX 2070 SUPER GPUs with an Intel Core i9-10900X CPU.
The last server had one NVIDIA RTX 3090 GPU with an AMD Ryzen 9 5950X CPU.
As we did not compare in terms of training speed, the trials were comparable across servers---DRL libraries use either TensorFlow or PyTorch, which conforms to 32-bit floating point precision by default.
Furthermore, when configuring the PPO implementations for RQ1 (and RQ2), we used a hyperparameter set similar to the one used by the original implementation~\cite{schulman2017proximal}, with the only difference being that we excluded LSTM layers in the neural network, to speed up training~\cite{shengyi2022the37implementation}.
We opted for the speed-up since this is a large-scale comparison-focused study, amounting to approximately 10K GPU hours.
When configuring Stable Baselines3's DDPG implementation for RQ3, we followed the hyperparameter set used by the replicated study~\cite{islam2017reproducibility}.
We trained five agents for each (1) (implementation, environment) permutation for RQ1 (and RQ2) and (2) (hyperparameter, value) permutation for RQ3. 
We trained five agents (\ie five trials) for each permutation as it was sufficient for reliable SBCI estimates with the Atari benchmark~\cite{agarwal2021deep}.

\begin{figure}[t]
    \centering
    \includegraphics[width=\linewidth]{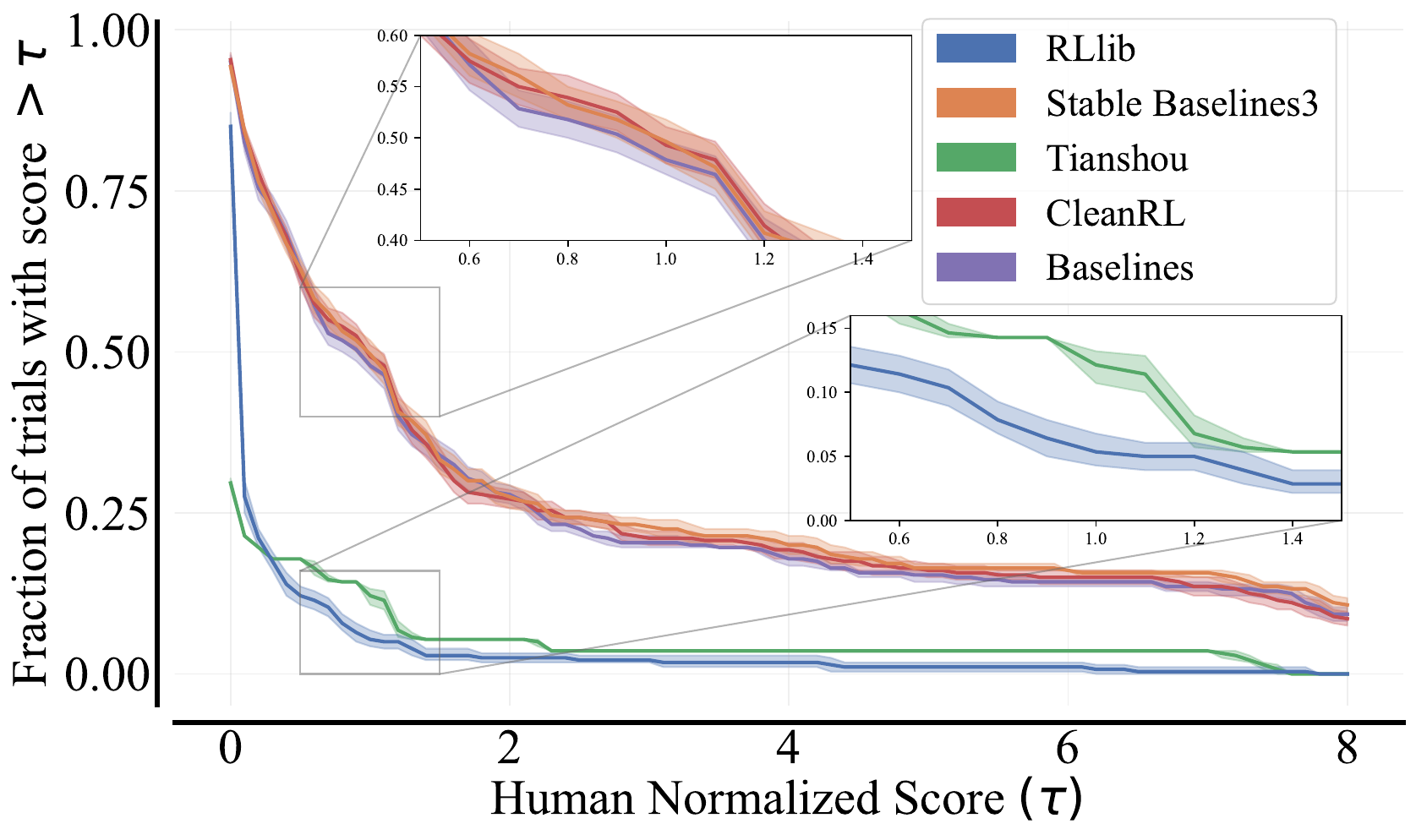}
    \caption{Performance profiles for the five PPO implementations tested across 56 environments, where the shaded regions indicate pointwise 95\% confidence bands based on SBCI.}
    \label{fig:diff_pp}
\end{figure}
\section{Results}

\begin{figure}[t]
    \centering
    \includegraphics[width=\linewidth]{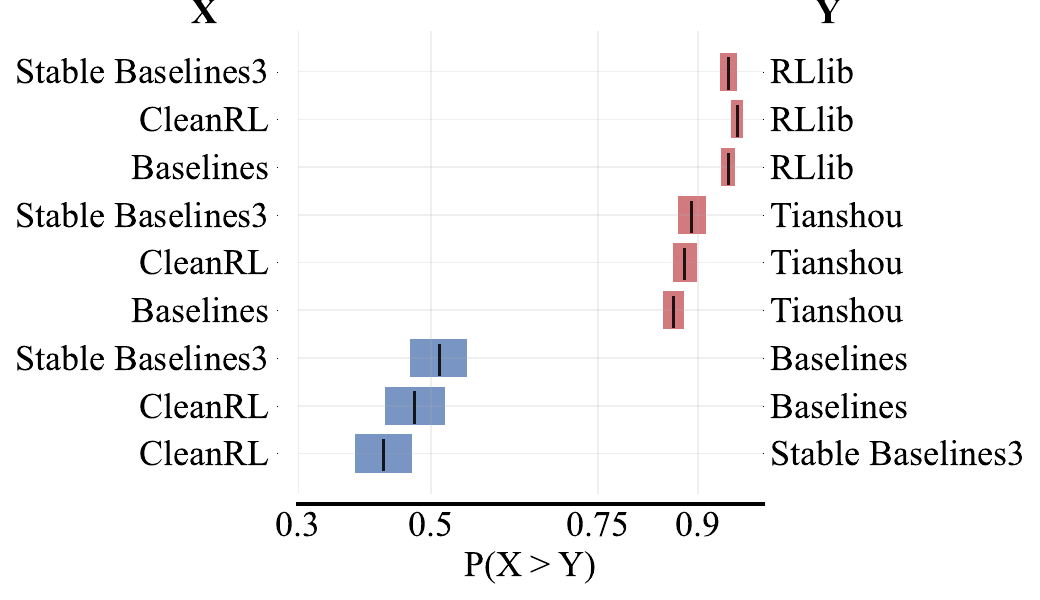}
    \caption{Pairwise POI for the five PPO implementations tested across 56 environments, where vertical stripes represent point estimates, and shaded regions indicate 95\% confidence bands based on SBCI. X is considered to be \emph{better} than Y if the point estimate is both statistically significant and meaningful. Comparisons where X is better than Y are indicated in red.}
    \label{fig:diff_poi}
\end{figure}
\begin{table*}[t]
    \centering
    \begingroup  
    \begin{threeparttable}
    \caption{Discrepancies Among The High-Performing PPO Implementations}
    \label{tab:diff_test_subset2}
        \begin{tabular}{@{}l@{}rrrrrrrcrrc@{}}
        \toprule
        \multirow{2.65}{*}{Game} & \multicolumn{4}{c}{Mean Reward Over 5 Trials} & & \multicolumn{6}{c}{One-way ANOVA\tnote{1}}\\
        \cmidrule{2-5} \cmidrule{7-12} 
        & Stable Baselines3 & CleanRL & Baselines & Baselines108 &  & F-statistic\tnote{2} & p-value\tnote{2} & Reject\tnote{2} &F-statistic\tnote{3} & p-value\tnote{3} & Reject\tnote{3} \\
        \midrule
        \begin{filecontents*}{csvs/diff_test_fixed.csv}
Game,Stable Baselines3,CleanRL,Baselines,F-statistic,p-value,p-value < 0.05,Baselines108K,F-statistic108K,p-value108K,p-value108K < 0.05
Atlantis,881191.6,946388.0,2088141.2,262.634,1.241e-10,Yes,848217.6,2.925,9.229e-02,No
ChopperCommand,5519.0,6056.8,841.4,157.884,2.408e-09,Yes,913.6,152.909,2.897e-09,Yes
DoubleDunk,-2.7,-3.4,-13.4,956.997,5.850e-14,Yes,-3.0,2.713,1.066e-01,No
Gopher,1521.1,4946.4,4752.1,8.636,4.747e-03,Yes,5086.4,8.189,5.718e-03,Yes
Krull,10369.7,9589.3,9535.5,17.274,2.936e-04,Yes,9653.7,4.453,3.577e-02,Yes
Robotank,19.8,9.5,13.9,5.365,2.165e-02,Yes,12.7,6.366,1.305e-02,Yes
Tennis,-4.2,-4.2,-14.2,7.101,9.229e-03,Yes,-3.9,0.004,9.959e-01,No
VideoPinball,37962.8,55677.5,48812.4,5.875,1.664e-02,Yes,72580.8,4.793,2.951e-02,Yes
Zaxxon,10241.2,5548.4,6024.6,6.263,1.372e-02,Yes,6685.6,7.351,8.238e-03,Yes
\end{filecontents*}
\csvreader[late after line=\\,late after last line=]{csvs/diff_test_fixed.csv}{}{\csvcoli & \csvcolii & \csvcoliii& \csvcoliv& \csvcolviii & & \csvcolv & \csvcolvi &  \ifthenelse{\equal{\csvcolvii}{Yes}}{\textbf{Yes}}{\csvcolvii} & \csvcolix& \csvcolx& \ifthenelse{\equal{\csvcolxi}{No}}{\textbf{No}}{\csvcolxi}}\\
        \bottomrule
        \end{tabular}
        \begin{tablenotes}
            \item[1] Reject null hypothesis of equal implementation means if p-value $<$ 0.05.
            \item[2] With respect to Stable Baselines3, CleanRL, and Baselines.
            \item[3] With respect to Stable Baselines3, CleanRL, and Baselines108---108K frames per episode.
        \end{tablenotes}
    \end{threeparttable}
    \endgroup
\end{table*}
\begin{figure*}[t]
    \centering
    \includegraphics[width=\linewidth]{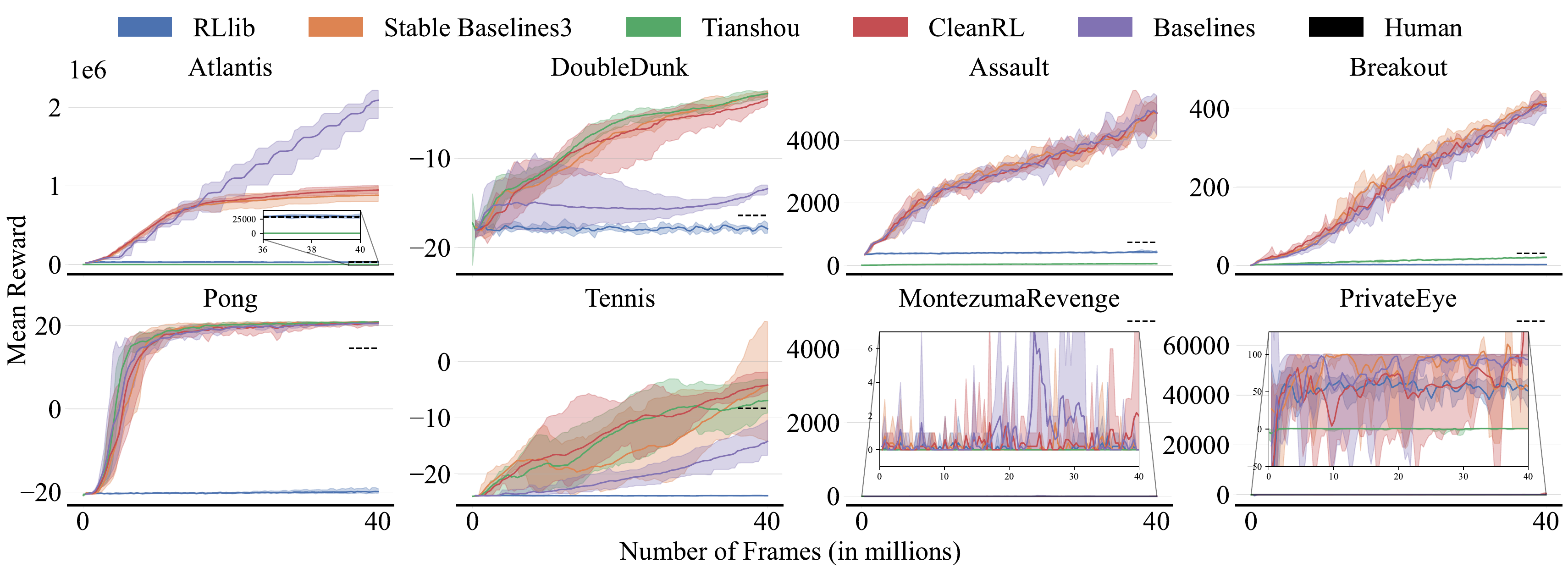}
    \caption{Training curves from five PPO implementations. Similar to DQN, five agents were trained for each (implementation, environment) permutation and the training curves were aggregated to display the mean, minimum, and maximum within the shaded regions. The mean reward attained by a professional human tester is also shown in black to gauge superhuman or subhuman capabilities. For brevity, only eight representative environments (out of 56) are displayed.}
    \label{fig:diff_subset}
\end{figure*}
In this section, we present the outcomes from our experiments addressing RQ1, RQ2, and RQ3.
\subsection{Prevalence of Implementation Discrepancies}
The performance profiles in~\autoref{fig:diff_pp} demonstrate stark discrepancies between the five PPO implementations tested.
Specifically, we can see two groups with similar performance; (1) Stable Baselines3, CleanRL, and Baselines where 50\% of their trials attained superhuman performance (\ie $\tau > 1$), and (2) RLlib and Tianshou where less than 15\% of their trials attained superhuman performance.
For conciseness, we subsequently refer to these two groups as high-performing and low-performing groups respectively. Although it cannot yet be inferred that the high-performing implementations are, in general, superior in terms of performance.

\paragraph{Probability of improvement between implementations}\autoref{fig:diff_poi} shows that the high-performing group is objectively better than the low-performing group. 
In particular, the pairwise POI between the two groups is both statistically significant, where $P(X>Y)>0.5 \; \wedge\; 0.5\notin CI$, and meaningful, where $CI_{upper}>0.75$.
From the pairwise comparisons, it is also observed that Stable Baselines3, CleanRL, and Baselines are on par with each other, with the pairwise comparisons among themselves being neither statistically significant nor meaningful.
However, as shown in~\autoref{tab:diff_test_subset2}, when considering individual environments, we found statistically significant discrepancies.
We later determined these discrepancies to be caused by implementation inconsistencies in RQ2.

\paragraph{Per-environment discrepancies} The one-way ANOVA in \autoref{tab:diff_test_subset2} with respect to Baselines (not the 108 variant) shows that the PPO implementations from the high-performing group were found to significantly differ in nine environments, contrasting the above-discussed aggregated comparisons.
This discrepancy can also be observed in the training curves, shown in~\autoref{fig:diff_subset}, with Atlantis and DoubleDunk. 
Specifically, the curves from Stable Baselines3 and CleanRL noticeably differ from Baselines' curves.

The PPO implementations from the low-performing group were observed to attain low mean rewards in the majority of environments, similar to the pattern observed in Assault and Breakout.
Tianshou could, however, solve some \emph{simple} environments, like Pong and Tennis.
These environments, however, have low score caps that are easily reached by agents~\cite{aitchison2023atari}, and therefore, not a strong indication of an implementation's predictive power.
Lastly, all five PPO implementations were unable to solve some environments like MontezumaRevenge and PrivateEye. 
This was, however, expected because these environments have sparse rewards~\cite{bellemare13arcade}, making them especially \emph{hard} in the context of DRL, where agents use rewards as feedback for previously executed actions.

\paragraph{Key takeaways}All of the aforementioned observations point strongly towards code-level inconsistencies and bugs between the tested PPO implementations.
In particular, environment-dependent inconsistencies between the high-performing implementations, and critical bugs with the low-performing implementations.
This is surprising, considering that (1) PPO is a common algorithm, (2) the implementations were from established libraries, and (3) Atari 2600, one of the most widely known benchmarks, was used.

Thus, researchers should exercise caution when interchanging implementations of the PPO algorithm as their experiments might yield different outcomes, depending on the implementation and environment used.
Furthermore, since code-level inconsistencies are observed to be a recurring pattern (also with DQN implementations), this now raises concerns about the reliability of the implementations of other less mature algorithms as well.
Researchers should not take the reliability and interchangeability of implementations for granted. 

\begin{tcolorbox}[sharp corners, colback=rqblue, colframe=rqblue]
    \textbf{RQ1 Summary:} Discrepancies between different implementations of the same algorithm are prevalent. In particular, out of the five popular PPO implementations tested on 56 environments, three implementations attained superhuman performance for 50\% of their trials while the other two implementations only attained superhuman performance for less than 15\% of their trials.
    Moreover, among the high-performing implementations, statistically significant performance differences were also observed in nine environments.
\end{tcolorbox}
\subsection{Reasons for Discrepancies}
\begin{figure}[t]
    \centering
    \includegraphics[width=\linewidth]{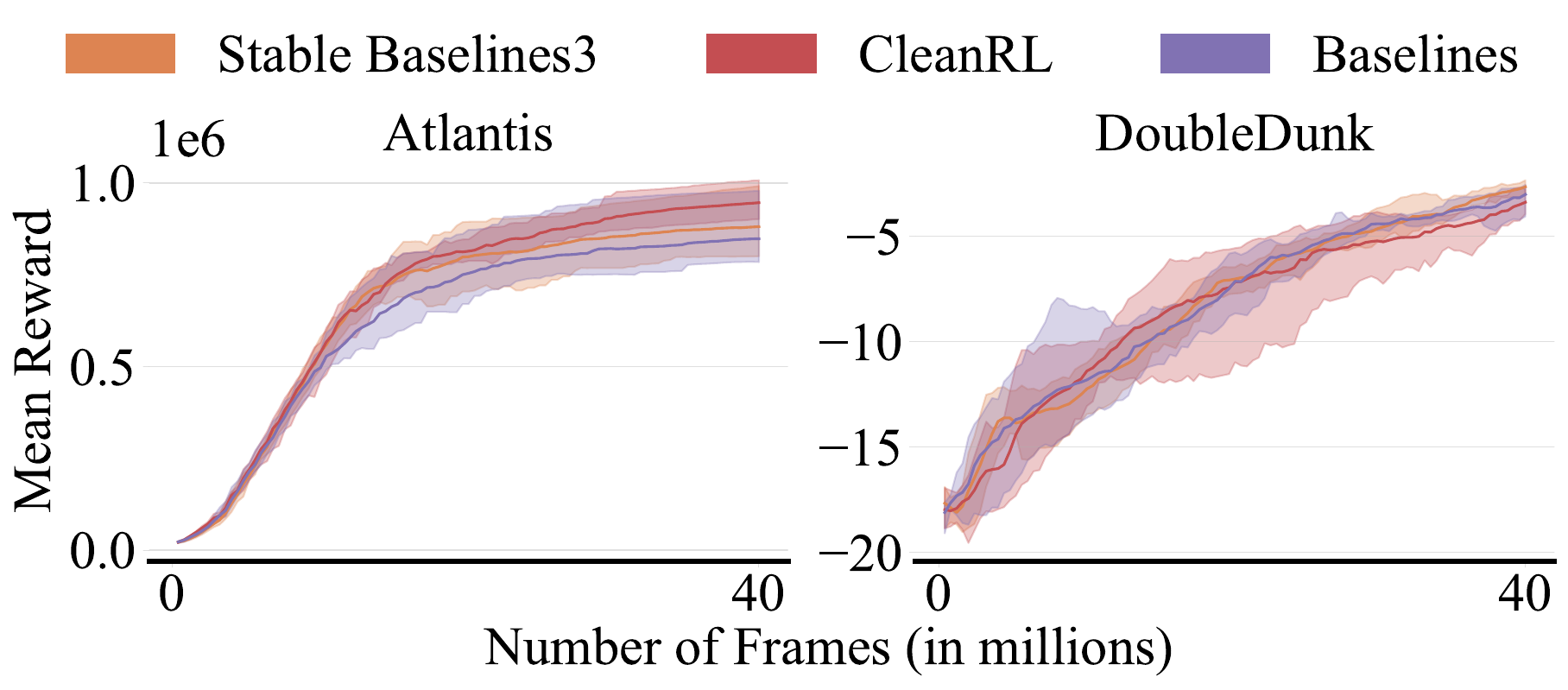}
    \caption{Consistent training curves among the high-performing PPO implementations after explicitly configuring the maximum frames per episode to be 108K. Like RQ1, each (implementation, environment) permutation underwent five trials.}
    \label{fig:debug}
\end{figure}
\begin{figure*}[t]
    \centering
    \includegraphics[width=\textwidth]{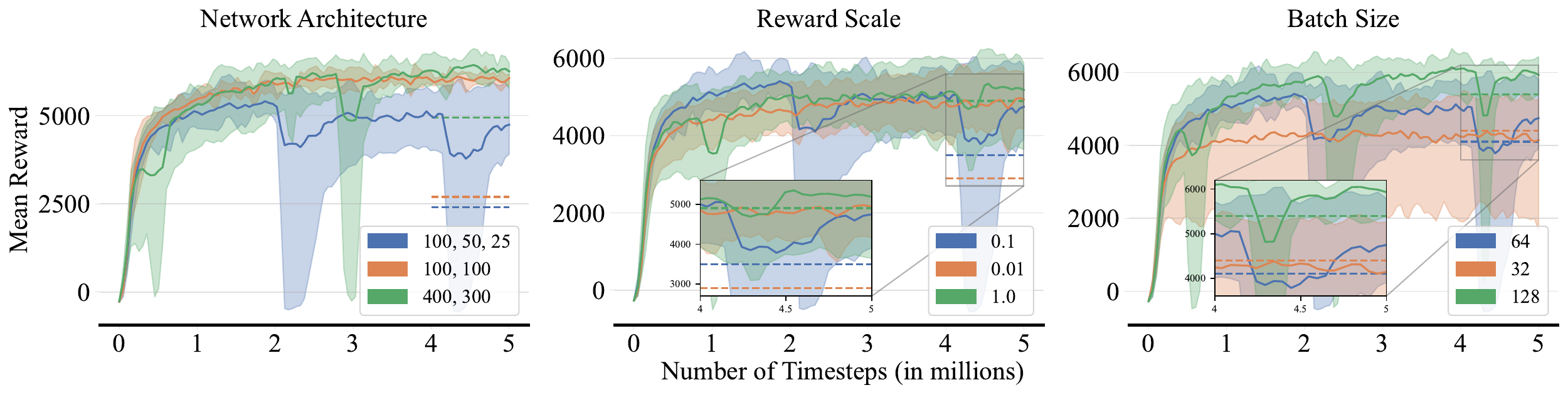}
    \caption{Training curves for three DDPG hyperparameter experiments on the HalfCheetah environment. Similar to PPO, five agents were trained for each (hyperparameter, value) permutation and the training curves were aggregated to display the mean, minimum, and maximum within the shaded regions. The dotted lines represent the efficacy reported in the replicated study.}
    \label{fig:rep_training_curves}
\end{figure*}
\begin{figure}[t]
    \centering
    \includegraphics[width=\linewidth]{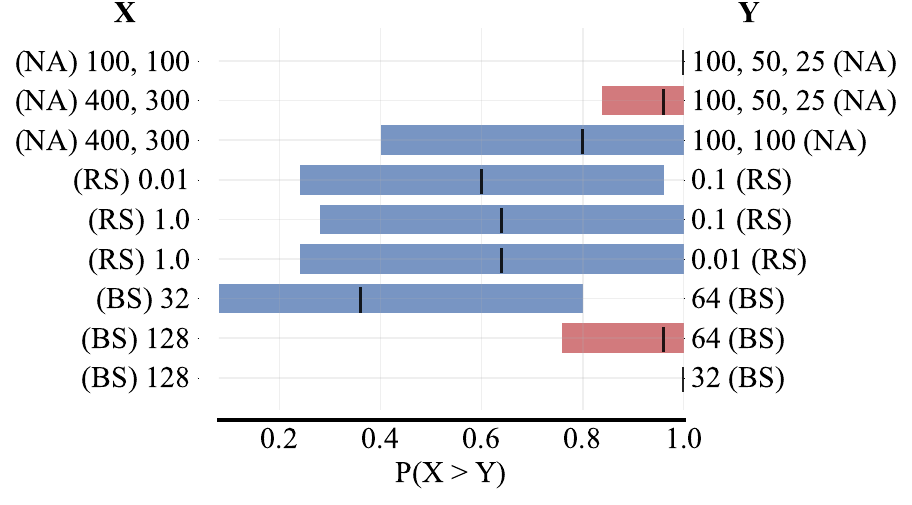}
    \caption{Pairwise POI for three DDPG hyperparameter experiments, where vertical stripes represent point estimates, and shaded regions represent 95\% confidence bands based on SBCI. X is considered to be \emph{better} than Y if the point estimate is both statistically significant and meaningful. Comparisons where X is better than Y are indicated in red.}
    \label{fig:rep_poi}
\end{figure}
In our manual analysis, we found three code-level inconsistencies among the high-performing PPO implementations, one of which explained the performance differences.
In particular, the inconsistency with the number of frames per episode influenced the performance while the inconsistencies with gradient clipping and averaging did not.

\paragraph{Frames per episode} Firstly, Baselines used the Gym library~\cite{gym} for the Atari 2600 benchmark, which defaulted to \emph{more than} 108K frames per episode while Stable Baselines3 and CleanRL used the Gymnasium library~\cite{gymnasium}, which defaulted \emph{strictly} to 108K frames per episode.
This was one of the reasons for the discrepancies observed in RQ1. 
Since the frames per episode were higher in Baselines, their agents spent more time in an environment before it gets reset---directly affecting the training process.
After correcting this by configuring \mintinline{python}{max_episode_steps} to be 27K (108K frames) in Baselines' environment preprocessing function \mintinline{python}{make_atari}, the implementations no longer significantly differed in three environments (out of nine), as seen in~\autoref{tab:diff_test_subset2}.
Furthermore, the training curves among the implementations were also observed to be more consistent when compared to those of RQ1, as shown in~\autoref{fig:debug}.

\paragraph{Gradient clipping} Secondly, the implementations differed in how they clipped gradients---commonly used to stabilize updates in a neural network and is dependent on the DL library used.
In particular, Stable Baselines3 and CleanRL used \mintinline{python}{clip_grad_norm} from the PyTorch library to clip gradients while Baselines used \mintinline{python}{clip_by_global_norm} from the TensorFlow library. 
These two functions clip the gradients differently.
To determine if this clipping inconsistency caused the discrepancies observed in RQ1, we (1) disabled gradient clipping across all high-performing implementations, and (2) re-tested and re-compared the implementations.
The performance without gradient clipping exhibited the same discrepancies as before and therefore, the clipping inconsistencies were not significant enough to have caused the discrepancies.

\paragraph{Gradient averaging} Lastly, Baselines also averaged gradients over multiple processes using OpenMPI~\cite{gabriel04:_open_mpi} to make updates in the neural network more efficient, something not implemented by either Stable Baselines3 or CleanRL.
As this would have affected the gradients in the neural network, we (1) disabled gradient averaging in Baselines, and (2) re-tested and re-compared the implementations.
The performance without gradient averaging exhibited the same discrepancies as before and therefore, the averaging inconsistencies were not significant enough to have caused the discrepancies.

\paragraph{Key takeaways} 
We found multiple implementation inconsistencies, one of which had a significant impact on performance.
The frames per episode inconsistency is an example of a difference that has a major impact on performance and that even the original implementation can be inconsistent. In fact, there are six environments in~\autoref{tab:diff_test_subset2} that still differ---hinting at more inconsistencies among the high-performing implementations.
The three inconsistencies observed can be classified as either API-based where the issue stems from the use of an external API (like frames per episode or gradient clipping), or logic-based where the issue lies internally (like gradient averaging)---types of inconsistencies also commonly found among AI implementations~\cite{chen2023toward,ahmed2023characterizing}. 
Moreover, we do not believe that these inconsistencies are in any way intentional, but rather, oversights stemming from a domain of high complexity, a phenomenon commonly observed with DL implementations~\cite{chen2023toward}.
This reaffirms the notion that it is \textit{not} necessarily the case that DRL implementations are consistent and interchangeable because of the \textit{high} risk of inconsistencies associated with the DRL paradigm---being a combination of both DL and RL. 

\begin{tcolorbox}[sharp corners, colback=rqblue, colframe=rqblue]
    \textbf{RQ2 Summary:} Implementation discrepancies occur primarily because of code-level inconsistencies between the implementations. In particular, we found three inconsistencies among the high-performing PPO implementations. Furthermore, correcting these inconsistencies solved discrepancies among the high-performing implementations in three environments.
\end{tcolorbox}

\subsection{Altering Experiment Outcomes}
The study we replicated explored the best values, in terms of efficacy, for commonly used hyperparameters with DDPG on the HalfCheetah environment.
Using a different DDPG implementation, we replicated the experiments for three hyperparameters, to wit, network architecture (NA), reward scale (RS), and batch size (BS).
The training curves in~\autoref{fig:rep_training_curves} and their respective pairwise POI in~\autoref{fig:rep_poi} contrast those originally reported by the replicated study.

\paragraph{Network architecture}
Firstly, for the experiment on network architecture, it can be seen in the training curves that the (400, 300) architecture variant does \emph{not} dominate the other two variants as previously reported in the study (\ie the dotted lines), but instead, is on par with the (100, 100) variant, and that both of them dominate the (100, 50, 25) variant. 
Their respective pairwise POI reaffirms this, where (1) both (400, 300) and (100, 100) are better than (100, 50, 25), and (2) (400, 300) is not better than (100, 100). 

\paragraph{Reward scale}
Secondly, for the experiment on reward scale, it can be seen in the training curves that a scale of 1 (\ie no scaling) does \emph{not} dominate the other two variants as previously reported in the study, but instead, is on par with both of them.
Their respective pairwise POI reaffirms this, where neither variant was better than the other.

\paragraph{Batch size}
Lastly, for the experiment on batch size, the training curves show that the 128 variant dominates the other two variants, similar to what was previously reported in the study.
Their pairwise POI reaffirms this, where (1) 128 is better than both 64 and 32, and (2) 32 is not better than 64.

\paragraph{Implications} We have demonstrated that the empirical outcomes originally reported by the study, although not incorrect, can differ based on the algorithm implementation used. Given that the study aimed to guide researchers towards ideal hyperparameter configurations, researchers could build on the conclusions of the study and \emph{not} see the claimed advantages due to these variances in outcomes.

\paragraph{Key takeaways}
This study was just \emph{one of the many} studies that we could have replicated and we selected it because it adheres to the current best evaluation practices.
However, the fact that using a different implementation was significant enough to alter experiment outcomes has serious implications on existing DRL research.
In particular, with research assuming that implementations are interchangeable~\cite{zeng2024offline,islam2017reproducibility,raghunath2022reinforcement,wolk2022beyond}, possibly warranting follow-up studies to replicate them and test whether their claims can be reproduced.

\begin{tcolorbox}[sharp corners, colback=rqblue, colframe=rqblue]
    \textbf{RQ3 Summary:} The assumption of interchangeable implementations was found to be significant enough to be able to alter the outcomes of an experiment. In particular, two out of three experiment outcomes we replicated with a different DDPG implementation differed from those of the original study.
\end{tcolorbox}

\section{Discussion}
In this section we discuss the (1) key takeaways, (2) actionable recommendations, and (3) issues and pull requests we filed for the discrepancies and inconsistencies found.

\paragraph{Implementations are not interchangeable}
In this study, we have shown that the assumption of interchangeable DRL implementations is (1) common among studies, (2) detrimental to a study's internal validity, potentially \textit{flipping} experiment outcomes, and thus, (3) \textit{mistaken}.
Hence, we recommend for replicability studies to test whether the claims of notable studies under this assumption can be reproduced.
Furthermore, we recommend for the replicability studies to adopt the differential testing methodology used in this study to increase the fairness of their comparisons.

\paragraph{Implementation inconsistencies exist}
In RQ1 and RQ2, we have shown that implementation discrepancies are prevalent and that code-level inconsistencies cause these discrepancies.
Two out of three inconsistencies found were API-based, suggesting that the complexity of DRL implementations, including their hyperparameters and stochasticity, make it difficult for unit testing to eliminate potential issues---something also observed in the DL domain~\cite{chen2023toward}. 
Thus, we believe that individually checking implementations will not solve the inconsistencies in the long run.
The inconsistencies can instead, be addressed with a more sustainable approach, at either the implementation or usage stage.

Firstly, for the implementation stage, we recommend that developers of DRL libraries proactively compare against other implementations using the differential testing methodology proposed in this study.
This differential testing methodology can be easily automated with CI/CD and is already incorporated in other domains to address implementation inconsistencies as well~\cite{diffwatch,louloudakis2023deltann}.
In fact, in response to our GitHub issue, Tianshou indicated that they already had plans for this.
Secondly, for the usage stage, we recommend that developers of DRL libraries explicitly document code-level inconsistencies for their implementations. Researchers should conduct their experiments with this in mind---acknowledging that these inconsistencies exist and avoiding the risk of basing their conclusions on them. 
We believe that this paves the way for more reliable and consistent implementations.

\paragraph{Testing granularities}
Although the differential testing methodology used in this study is end-to-end (\ie testing the entire algorithm), we would like to highlight that differential testing of individual functions or components is feasible as well.
However, one significant drawback in doing so is that only a subset of functions can be tested, as noted by Herbold and Tunkel~\cite{herbold2023differential} when they differential tested \textit{Machine Learning} (ML) libraries.
This is because of differences between the hyperparameters and implementations of the functions. For example, some DRL libraries might expose only a subset of hyperparameters in their functions, or even combine functions together, making it difficult to differential test them.
Thus, we opted to test end-to-end for completeness.
Nonetheless, functional testing has its advantages as well; (1) computational costs can be reduced by testing only what needs to be tested, and (2) inconsistencies can be localized more easily by only inspecting the functions that were tested.
Thus, we recommend for researchers and practitioners to test end-to-end when prioritizing completeness and at a functional level when prioritizing computational costs and inconsistency localization.

\paragraph{Large test suites are preferred}
A test suite's size is correlated with its effectiveness~\cite{kochhar2015code} and should cover as many situations as possible to increase its effectiveness~\cite{namin2009influence}.
Had we instead opted for any of the \emph{optimal} Atari environment subsets proposed by Atari-5~\cite{aitchison2023atari}, discrepancies among the high-performing PPO implementations with six environments in~\autoref{tab:diff_test_subset2} would still have gone \emph{undetected}.
This underscores the importance of using a large environment suite for comparative studies in the DRL domain.
This is even more applicable to studies on the extent of implementation discrepancies, where existing studies in this area only used a small environment suite for their experiments~\cite{Engstrom2020Implementation,henderson2018deep}.
Thus, we recommend for DRL researchers and practitioners to prioritize using a large environment suite whenever possible, with SBCI.
The additional computation costs can be significantly reduced by using SBCI to effectively minimize the number of trials required for accurate and reliable estimates.

\paragraph{Issues and pull requests}
In total, we filed six issues and one pull request on GitHub; (1) five issues and one pull request for the discrepancies and inconsistencies found in RQ1 and RQ2---one issue for each DRL library tested and one pull request for Baselines' \textit{frames per episode} inconsistency, and (2) one issue for a discrepancy we found with RLlib's SAC~\cite{christodoulou2019soft} implementation during our preliminary experiments.
In response to our issues, the developers from Stable Baselines3 and CleanRL uncovered two additional inconsistencies regarding timeouts and value clipping that could account for the discrepancies among the remaining six environments in~\autoref{tab:diff_test_subset2}.
The developers from Tianshou actively investigated and confirmed that the cause for the discrepancies was because of a misleading code example in their repository while the developers from Baselines and RLlib have yet to acknowledge our issues or pull request.
RLlib's developers, however, acknowledged our issue regarding the SAC discrepancy, which was later confirmed to be because of an inconsistency.

\section{Threats To Validity}
Here, we detail the steps taken to assess and mitigate the most important threats to validity.

\paragraph{Internal validity} Firstly, one concern was with incorrectly configuring the implementations.
Since libraries aim to implement algorithms that are both logically the same and compatible with their existing architecture (\eg by inheriting from existing classes core to the library), they end up having different naming conventions and function signatures.
For example, all of the PPO code for CleanRL is contained within a single file with minimal functions, while the PPO code for RLlib spans multiple files and functions.
This diversity among implementations is not specific to DRL and is also prevalent in other domains as well, like ML~\cite{herbold2023differential}.

To reduce the likelihood of a configuration oversight, we closely followed the code examples provided by the implementations, modifying only where necessary.
Furthermore, to ascertain that the discrepancies found were not from a configuration oversight, we adopted a \emph{best effort} approach when debugging them.
In particular, we (1) manually re-inspected the code for inconsistencies that could account for the discrepancies, (2) re-tested and re-compared when an inconsistency was found, and (3) contacted the implementations' developers when we could not fix the discrepancy, as a final attempt to address it.
Moreover, we focused on the high-performing implementations as their discrepancies were more likely to stem from actual implementation inconsistencies, rather than configuration oversights or critical bugs.
Nonetheless, we still informed developers from all DRL libraries of the discrepancies found.
Lastly, we are confident that any configuration oversights would not affect the key takeaways of this study, as they could at most affect a small subset of the results—--the discrepancies without known inconsistencies from RLlib's PPO implementation.

\paragraph{External validity} Secondly, one other concern is that only implementations of two algorithms were tested.
We selected DQN and PPO because they are both common (implemented in 16 out of the 18 DRL libraries inspected) and from different DRL paradigms (QL and PG).
If discrepancies and code-level inconsistencies were prevalent among their implementations, there is a high likelihood that they are prevalent among implementations of other algorithms as well. For example, inconsistencies we discovered \emph{a year ago} with RLlib's implementation of SAC with their 2.2.0 release that are still present in their current release---2.34.0.
Furthermore, their experimental results could potentially be extrapolated to other similar algorithms in their respective paradigms as well, as demonstrated with DDPG in RQ3.

\section{Related Work}
In this section, we discuss existing related literature.
\paragraph{Implementation discrepancies in DRL}
While there have been recent studies reporting implementation discrepancies in DRL, it was not their primary focus. 
One study focused on the nuances between two \emph{different} algorithms~\cite{Engstrom2020Implementation} while another focused on how different values for hyperparameters influenced the outcome~\cite{henderson2018deep}, similar to~\cite{islam2017reproducibility}---the study we replicated in RQ3.
Building on the foundations laid by these studies, our work incorporates several additional methodologies to further deepen the understanding of implementation discrepancies; (1) \textit{systematic} literature reviews and implementation selection, (2) \textit{large} test suites, (3) DRL tailored \textit{statistical} techniques, and (4) root cause \textit{investigation}.

\paragraph{Differential testing in AI}
There has been an abundance of studies using differential testing to identify and localize discrepancies in AI.
In DL, the most common approaches include either fuzzing, mutation, or targeted searching to generate test cases for differential testing~\cite{deng2022fuzzing,guo2020audee,wei2022free,duo,fuzzgpt}, and can range from applications such as generating adversarial inputs~\cite{guo2018dlfuzz,pei2017deepxplore,riccio2020model}, testing DL implementations~\cite{wang2022eagle,deng2023differential}, and large language model based fuzzers~\cite{fuzz4all,titanfuzz,yang2023white,yang2023kernelgpt}.
Moreover, there have also been comparative studies similar to this study, but instead, accessing DL and ML implementations~\cite{dai2022reveal,novac2022analysis,herbold2023differential}.
Different from these studies, we now apply differential testing in the DRL domain to investigate the assumption of interchangeable DRL implementations.

\section{Conclusion}
We conducted a large-scale testing-focused study to investigate the assumption of interchangeable DRL implementations.
We observed significant discrepancies among the different implementations, later determined to be caused by code-level inconsistencies.
We demonstrated that this assumption of implementation interchangeability was then significant enough to alter experimental outcomes.
Lastly, we provided recommendations for DRL researchers and practitioners on how to reliably use, compare, and test DRL implementations.
With the emerging field of integrating DRL with software testing~\cite{ahmad2020using,nouwou2023comparison,kim2018generating}, it is imperative to assess and improve the reliability of DRL implementations and studies, sooner rather than later.

\section*{Acknowledgments}
This research was supported by a Ministry of Education (MOE) Academic Research Fund (AcRF) Tier 1 grant and the Shenzhen Science and Technology Program (No. KJZD20240903095700001).

\balance
\bibliographystyle{IEEEtran}
\bibliography{references}

\onecolumn
\newpage
\renewcommand{\appendixname}{Supplementary Material}
\appendix
\setcounter{table}{0}
\setcounter{figure}{0}

\section{PPO Trials}
\begin{table*}[h]
    \centering
    \begingroup  
    \begin{threeparttable}
    \caption{Discrepancies Among All Five PPO Implementations}
    \label{tab:supp1}
        \begin{tabular}{@{}l@{\qquad\qquad}rrrrrrrrc@{}}
        \toprule
        \multirow{2.65}{*}{Game} & \multicolumn{5}{c}{Mean Reward Over 5 Trials} & 
 & \multicolumn{3}{c}{One-way ANOVA\tnote{1}}\\
        \cmidrule{2-6} \cmidrule{8-10} 
        & RLlib & Stable Baselines3 & Tianshou & CleanRL & Baselines & & F-statistic & p-value & Reject\\
        \midrule
        \begin{filecontents*}{csvs/diff_test.csv}
Game,RLlib,Stable Baselines3,Tianshou,CleanRL,Baselines,F-statistic,p-value,p-value < 0.05
Alien,290.4,1395.0,33.4,1609.1,1751.6,36.042,7.044e-09,Yes
Amidar,71.3,581.2,19.7,628.3,628.9,74.773,9.706e-12,Yes
Assault,421.4,4888.5,51.0,4867.6,4880.6,273.808,3.723e-17,Yes
Asterix,301.3,4031.1,22.1,3450.6,4206.4,63.464,4.432e-11,Yes
Asteroids,1239.4,1637.1,8.2,1476.3,1519.6,169.69,3.953e-15,Yes
Atlantis,29882.2,881191.6,632.0,946388.0,2088141.2,692.892,3.894e-21,Yes
BankHeist,38.1,700.3,30.1,1124.4,1002.5,17.32,2.788e-06,Yes
BattleZone,3486.0,23426.0,3.9,19256.0,20734.0,41.093,2.237e-09,Yes
BeamRider,466.6,3255.2,22.2,2350.8,2631.7,42.124,1.797e-09,Yes
Berzerk,670.7,1011.5,6.3,928.8,1124.0,98.94,6.980e-13,Yes
Bowling,23.8,60.0,10.0,52.9,49.4,26.465,9.662e-08,Yes
Boxing,1.9,93.6,89.6,93.3,93.9,11707.755,2.210e-33,Yes
Breakout,1.9,418.4,20.4,410.6,407.2,2869.62,2.784e-27,Yes
Centipede,2640.1,3520.4,33.6,3704.0,3350.2,188.856,1.400e-15,Yes
ChopperCommand,1011.2,5519.0,13.6,6056.8,841.4,260.782,6.003e-17,Yes
CrazyClimber,27593.2,112393.2,124.4,109474.0,112234.0,466.387,1.964e-19,Yes
Defender,4598.3,52020.2,34.8,46683.3,47417.7,388.984,1.178e-18,Yes
DemonAttack,189.6,15142.2,28.8,13683.0,16628.4,102.111,5.175e-13,Yes
DoubleDunk,-17.9,-2.7,-2.7,-3.4,-13.4,1606.585,9.064e-25,Yes
Enduro,0.0,881.9,939.6,1018.5,998.6,122.937,8.819e-14,Yes
FishingDerby,-88.6,18.0,27.2,24.7,24.4,113.104,1.956e-13,Yes
Freeway,12.8,33.1,33.0,33.0,33.1,14.935,8.368e-06,Yes
Frostbite,212.4,307.1,13.7,285.5,814.3,1.738,1.813e-01,No
Gopher,387.6,1521.1,38.0,4946.4,4752.1,21.752,4.749e-07,Yes
Gravitar,259.0,646.7,0.7,634.8,803.7,15.457,6.508e-06,Yes
Hero,1088.3,29961.0,43.6,28057.5,29299.8,183.683,1.833e-15,Yes
IceHockey,-6.3,-4.0,-4.3,-4.5,-4.5,16.507,4.002e-06,Yes
Jamesbond,38.8,555.2,1.7,504.9,537.6,103.011,4.762e-13,Yes
Kangaroo,90.4,8087.0,14.2,4623.0,4368.0,4.914,6.343e-03,Yes
Krull,3881.9,10369.7,234.4,9589.3,9535.5,222.1,2.885e-16,Yes
KungFuMaster,13674.8,29643.0,42.5,28254.0,30090.0,24.902,1.594e-07,Yes
MontezumaRevenge,0.0,0.0,0.0,2.0,0.0,1.0,4.307e-01,No
MsPacman,926.8,2126.2,43.6,2248.3,2299.5,112.383,2.079e-13,Yes
NameThisGame,2384.6,5818.8,113.0,5822.7,6347.5,161.58,6.351e-15,Yes
Phoenix,878.9,10597.8,18.4,11236.1,10005.5,67.666,2.453e-11,Yes
Pitfall,-10.0,-4.7,-0.1,-1.2,-6.6,1.273,3.136e-01,No
Pong,-19.9,20.9,20.8,20.6,20.5,16732.836,6.223e-35,Yes
PrivateEye,51.0,56.4,0.8,264.9,93.0,1.824,1.638e-01,No
Qbert,437.2,15322.4,53.9,15838.8,16193.6,1091.993,4.239e-23,Yes
Riverraid,2477.7,8911.5,26.5,8207.9,9057.1,167.778,4.412e-15,Yes
RoadRunner,7159.8,34385.8,43.5,29516.8,43170.2,22.779,3.284e-07,Yes
Robotank,2.2,19.8,4.4,9.5,13.9,16.922,3.322e-06,Yes
Seaquest,498.6,1120.9,23.8,1565.4,1698.4,21.806,4.657e-07,Yes
Skiing,-16443.4,-30000.0,-2911.7,-30000.0,-23134.0,32.038,1.940e-08,Yes
Solaris,2278.0,2288.2,3.8,2514.9,2434.1,96.434,8.899e-13,Yes
SpaceInvaders,264.6,1138.4,26.3,1006.1,1105.0,54.866,1.679e-10,Yes
StarGunner,1091.2,38492.4,20.6,43477.2,39900.2,207.069,5.714e-16,Yes
Tennis,-23.9,-4.2,-6.9,-4.2,-14.2,22.987,3.051e-07,Yes
TimePilot,3502.6,6725.4,4.8,5974.2,6493.2,179.287,2.319e-15,Yes
Tutankham,122.8,241.3,34.1,230.6,211.2,161.039,6.560e-15,Yes
UpNDown,1855.9,203052.9,184.9,141372.4,224597.5,8.184,4.437e-04,Yes
Venture,0.0,20.4,0.1,19.8,9.0,0.567,6.897e-01,No
VideoPinball,24055.0,37962.8,54.1,55677.5,48812.4,57.966,1.017e-10,Yes
WizardOfWor,587.0,6355.4,7.4,7088.0,6197.0,59.251,8.326e-11,Yes
YarsRevenge,8497.5,22768.9,13.4,43311.5,26453.1,3.88,1.720e-02,Yes
Zaxxon,176.0,10241.2,3.6,5548.4,6024.6,29.353,4.073e-08,Yes
\end{filecontents*}
\csvreader[late after line=\\,late after last line=]{csvs/diff_test.csv}{}{\csvcoli & \csvcolii & \csvcoliii& \csvcoliv&\csvcolv & \csvcolvi & & \csvcolvii & \csvcolviii & \ifthenelse{\equal{\csvcolix}{Yes}}{\textbf{Yes}}{\csvcolix}}\\
        \bottomrule
        \end{tabular}
        \begin{tablenotes}
            \item[1] Reject null hypothesis of equal implementation means if p-value $<$ 0.05.
        \end{tablenotes}
    \end{threeparttable}
    \endgroup
\end{table*}
\begin{table*}[t]
    \centering
    \begingroup  
    \begin{threeparttable}
    \caption{Discrepancies Among The High-Performing PPO Implementations}
    \label{tab:supp2}
        \begin{tabular}{@{}l@{\qquad\qquad}rrrrrrc@{}}
        \toprule
        \multirow{2.65}{*}{Game} & \multicolumn{3}{c}{Mean Reward Over 5 Trials} & 
 & \multicolumn{3}{c}{One-way ANOVA\tnote{1}}\\
        \cmidrule{2-4} \cmidrule{6-8} 
        & Stable Baselines3 & CleanRL & Baselines & & F-statistic & p-value & Reject\\
        \midrule
        \begin{filecontents*}{csvs/diff_test_subset.csv}
Game,Stable Baselines3,CleanRL,Baselines,F-statistic,p-value,p-value < 0.05
Alien,1395.0,1609.1,1751.6,1.118,3.587e-01,No
Amidar,581.2,628.3,628.9,0.35,7.115e-01,No
Assault,4888.5,4867.6,4880.6,0.003,9.972e-01,No
Asterix,4031.1,3450.6,4206.4,1.405,2.830e-01,No
Asteroids,1637.1,1476.3,1519.6,1.729,2.188e-01,No
Atlantis,881191.6,946388.0,2088141.2,262.634,1.241e-10,Yes
BankHeist,700.3,1124.4,1002.5,1.838,2.012e-01,No
BattleZone,23426.0,19256.0,20734.0,0.949,4.145e-01,No
BeamRider,3255.2,2350.8,2631.7,2.698,1.078e-01,No
Berzerk,1011.5,928.8,1124.0,2.873,9.562e-02,No
Bowling,60.0,52.9,49.4,1.034,3.854e-01,No
Boxing,93.6,93.3,93.9,0.405,6.758e-01,No
Breakout,418.4,410.6,407.2,1.201,3.347e-01,No
Centipede,3520.4,3704.0,3350.2,1.748,2.157e-01,No
ChopperCommand,5519.0,6056.8,841.4,157.884,2.408e-09,Yes
CrazyClimber,112393.2,109474.0,112234.0,0.502,6.173e-01,No
Defender,52020.2,46683.3,47417.7,3.029,8.610e-02,No
DemonAttack,15142.2,13683.0,16628.4,1.927,1.881e-01,No
DoubleDunk,-2.7,-3.4,-13.4,956.997,5.850e-14,Yes
Enduro,881.9,1018.5,998.6,2.661,1.105e-01,No
FishingDerby,18.0,24.7,24.4,0.395,6.822e-01,No
Freeway,33.1,33.0,33.1,0.332,7.239e-01,No
Frostbite,307.1,285.5,814.3,1.065,3.751e-01,No
Gopher,1521.1,4946.4,4752.1,8.636,4.747e-03,Yes
Gravitar,646.7,634.8,803.7,0.765,4.869e-01,No
Hero,29961.0,28057.5,29299.8,0.453,6.459e-01,No
IceHockey,-4.0,-4.5,-4.5,1.784,2.097e-01,No
Jamesbond,555.2,504.9,537.6,0.509,6.133e-01,No
Kangaroo,8087.0,4623.0,4368.0,1.087,3.681e-01,No
Krull,10369.7,9589.3,9535.5,17.274,2.936e-04,Yes
KungFuMaster,29643.0,28254.0,30090.0,0.09,9.143e-01,No
MontezumaRevenge,0.0,2.0,0.0,1.0,3.966e-01,No
MsPacman,2126.2,2248.3,2299.5,0.819,4.639e-01,No
NameThisGame,5818.8,5822.7,6347.5,1.215,3.307e-01,No
Phoenix,10597.8,11236.1,10005.5,0.495,6.215e-01,No
Pitfall,-4.7,-1.2,-6.6,0.358,7.062e-01,No
Pong,20.9,20.6,20.5,2.182,1.555e-01,No
PrivateEye,56.4,264.9,93.0,1.321,3.031e-01,No
Qbert,15322.4,15838.8,16193.6,1.741,2.169e-01,No
Riverraid,8911.5,8207.9,9057.1,1.194,3.365e-01,No
RoadRunner,34385.8,29516.8,43170.2,2.065,1.696e-01,No
Robotank,19.8,9.5,13.9,5.365,2.165e-02,Yes
Seaquest,1120.9,1565.4,1698.4,2.651,1.113e-01,No
Skiing,-30000.0,-30000.0,-23134.0,2.474,1.260e-01,No
Solaris,2288.2,2514.9,2434.1,0.679,5.253e-01,No
SpaceInvaders,1138.4,1006.1,1105.0,0.587,5.710e-01,No
StarGunner,38492.4,43477.2,39900.2,1.695,2.248e-01,No
Tennis,-4.2,-4.2,-14.2,7.101,9.229e-03,Yes
TimePilot,6725.4,5974.2,6493.2,2.375,1.352e-01,No
Tutankham,241.3,230.6,211.2,2.939,9.148e-02,No
UpNDown,203052.9,141372.4,224597.5,0.789,4.765e-01,No
Venture,20.4,19.8,9.0,0.139,8.717e-01,No
VideoPinball,37962.8,55677.5,48812.4,5.875,1.664e-02,Yes
WizardOfWor,6355.4,7088.0,6197.0,0.677,5.265e-01,No
YarsRevenge,22768.9,43311.5,26453.1,0.994,3.985e-01,No
Zaxxon,10241.2,5548.4,6024.6,6.263,1.372e-02,Yes
\end{filecontents*}
\csvreader[late after line=\\,late after last line=]{csvs/diff_test_subset.csv}{}{\csvcoli & \csvcolii & \csvcoliii& \csvcoliv& &\csvcolv & \csvcolvi & \ifthenelse{\equal{\csvcolvii}{Yes}}{\textbf{Yes}}{\csvcolvii}}\\
        \bottomrule
        \end{tabular}
        \begin{tablenotes}
            \item[1] Reject null hypothesis of equal implementation means if p-value $<$ 0.05.
        \end{tablenotes}
    \end{threeparttable}
    \endgroup
\end{table*}
\begin{figure*}[t]
    \centering
    \includegraphics[width=\textwidth]{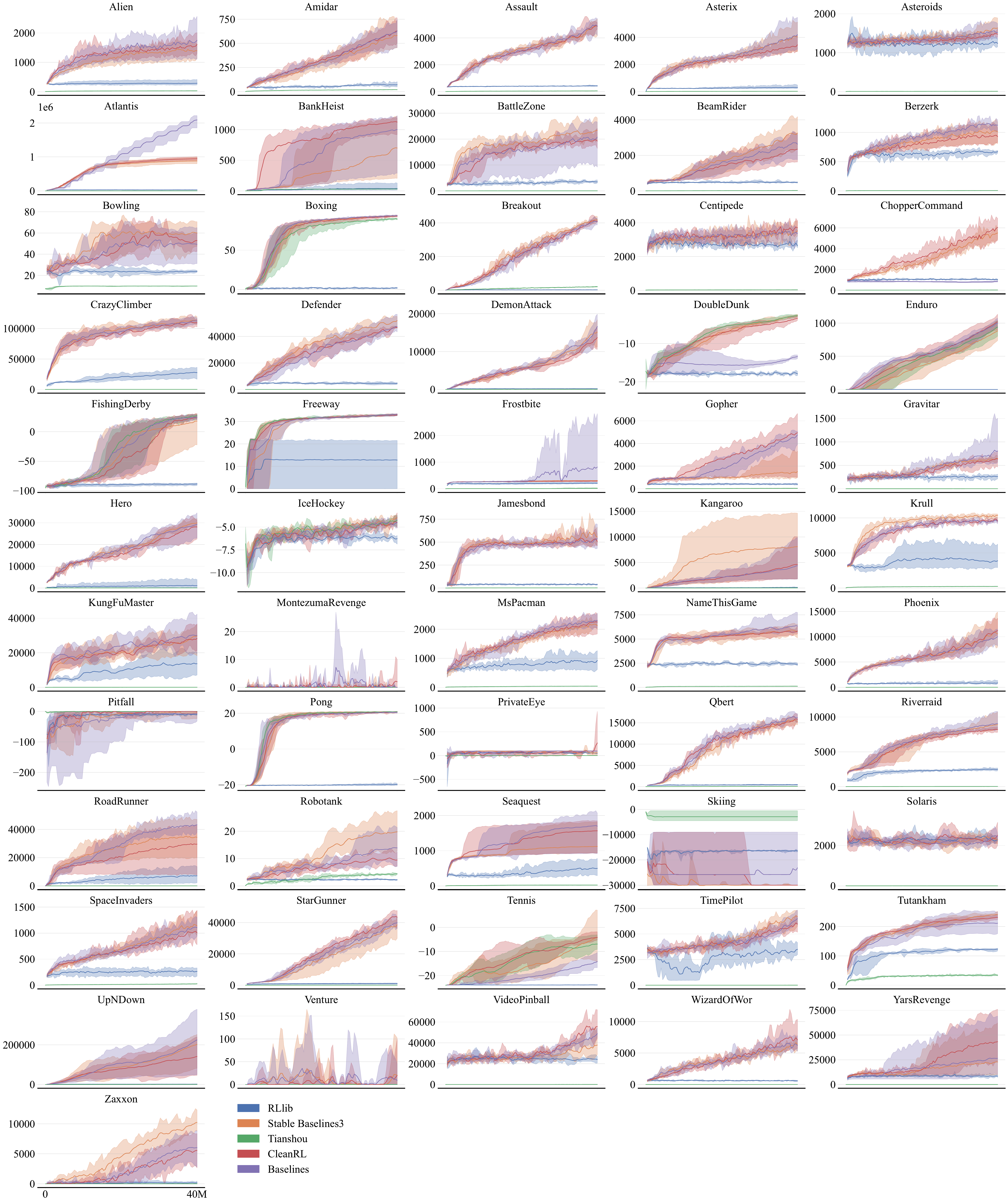}
    \caption{Training curves from five PPO implementations across all 56 environments where the y-axis and x-axis represent the mean reward and number of frames respectively. Five agents were trained for each (implementation, environment) permutation and the training curves were aggregated to display the mean, minimum, and maximum within the shaded regions.}
    \label{fig:supp1}
\end{figure*}
\clearpage
\begin{figure*}[ht!]
    \centering
    \includegraphics[width=\textwidth]{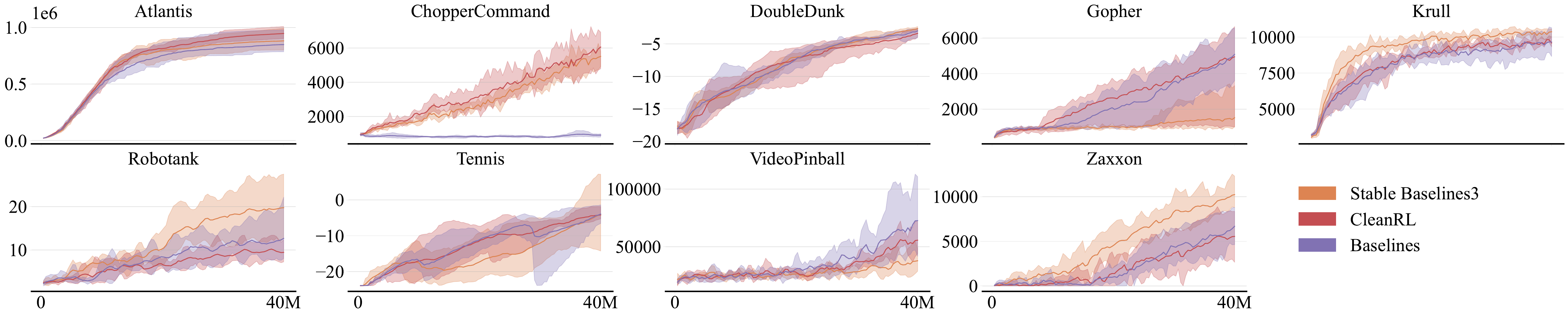}
    \caption{Training curves from the high-performing PPO implementations across the nine environments they significantly differed in (\emph{after} fixing the frames per episode inconsistency). The curves from ChopperCommand, Gopher, Robotank, and Zaxxon suggest that multiple undiscovered inconsistencies still exist between the implementations.}
    \label{fig:supp2}
\end{figure*}
\begin{figure*}[ht!]
    \centering
    \includegraphics[width=\textwidth]{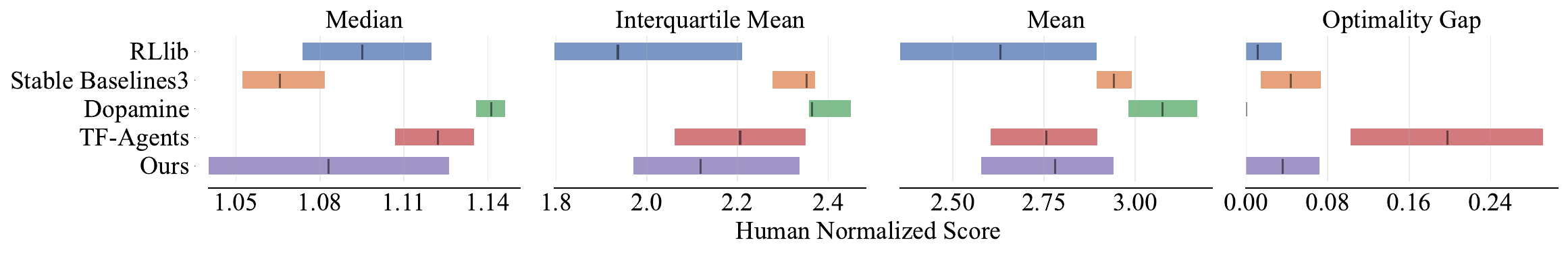}
    \caption{Applying SBCI to the trials from the \emph{pilot study} across four different aggregation metrics. Interquartile mean removes the highest and lowest 25\% from the distribution for a more robust estimate. Optimality gap measures the amount by which the trial fails to meet human-level performance---1.0. Hence, unlike other metrics, a lower optimality gap suggests better performance. The implementations are seen to vary greatly, in terms of performance, across all four aggregation metrics used.}
    \label{fig:supp3}
\end{figure*}
\end{document}